\documentclass[a4paper,fleqn]{cas-sc}

\usepackage[numbers]{natbib}
\usepackage{enumitem}  
\usepackage{float}
\usepackage{graphicx}
\usepackage{tabularx}
\usepackage{ltablex}     
\usepackage{multirow}
\usepackage{longtable}
\usepackage{fontawesome5}
\usepackage{array}
\usepackage{makecell}
\usepackage{placeins}

\newcolumntype{P}[1]{>{\centering\arraybackslash}p{#1}}

\usepackage{booktabs}    
\usepackage{fontawesome5} 
\usepackage{pdflscape}
\usepackage{listings}
\usepackage{xcolor}

\definecolor{darkgray}{gray}{0.4}
\definecolor{lightgray}{gray}{0.75}

\newcommand{\mybarhhigh}[2]{%
  {\color{black}\rule{#1mm}{4pt}}\hspace{2pt}#2%
}

\usepackage{graphicx}
\usepackage{subcaption}
\usepackage{calc}

\setlist[enumerate]{noitemsep, topsep=0pt}
\setlist[itemize]{noitemsep, topsep=0pt}

\def\tsc#1{\csdef{#1}{\textsc{\lowercase{#1}}\xspace}}
\tsc{WGM}
\tsc{QE}
\tsc{EP}
\tsc{PMS}
\tsc{BEC}
\tsc{DE}

\begin{document}
\let\WriteBookmarks\relax
\def\floatpagepagefraction{1}
\def\textpagefraction{.001}

\shorttitle{Engineering for Crisis Management}


\title [mode = title]{Engineering for Crisis Management: A User-Centred Analysis of Disaster Mobile Applications}                      



\author[1]{Muhamad Syukron}
\cormark[1]
\ead{muhamad.syukron@its.ac.id}

\author[2]{Anuradha Madugalla}
\author[3]{Mojtaba Shahind}
\author[4]{John Grundy}

\affiliation[1]{organization={Institut Teknologi Sepuluh Nopember},
    city={Surabaya},
    country={Indonesia}}

\affiliation[2]{organization={Deakin University},
    city={Melbourne},
    country={Australia}}

\affiliation[3]{organization={RMIT University},
    city={Melbourne},
    country={Australia}}

\affiliation[4]{organization={Monash University},
    city={Melbourne},
    country={Australia}}

\cortext[1]{Corresponding author}










\begin{abstract}
Disaster mobile apps play an increasingly important role in disseminating hazard information and supporting communities during emergency situations. This study presents a comprehensive analysis of these mobile applications, focusing on their features, user-reported challenges, and opportunities for improvement. We first examined the landscape of disaster mobile apps by analysing 70 apps identified through a combination of methods, including those from the literature, the Google Play Store, and the App Store. The analysis categorised apps based on disaster focus, geographic coverage, popularity, monetisation strategies, and features across the disaster lifecycle. We then extracted, translated and analysed user reviews using topic modelling and sentiment analysis to identify key concerns and recurring issues. The results show that most applications prioritise response-related functionalities, with limited support for preparedness and recovery. User feedback highlights critical challenges related to technical reliability, usability, accessibility, and information clarity. Based on these findings, we propose a set of recommendations for developers and emergency management agencies to improve the reliability, inclusiveness, and overall effectiveness of disaster mobile apps. These include adopting lifecycle-oriented design approaches, strengthening multilingual support, improving technical robustness, and integrating user feedback into development processes. This work contributes to the growing body of research on human-centred disaster risk reduction by providing empirical insights and actionable guidance for the design of more reliable and inclusive disaster communication systems.
\end{abstract}

\begin{keywords}
Mobile Apps \sep Safety-Critical Software \sep Software Engineering \sep Feature Analysis \sep Inclusivity
\end{keywords}

\maketitle

\section{Introduction}
Communities worldwide are facing increasingly frequent natural hazards, including bushfires, floods, cyclones, heatwaves, earthquakes and many more. As climate change accelerates the frequency of these hazards, timely risk information communication has never been more critical. This need is strongly emphasised in the Sendai Framework for Disaster Risk Reduction 2015–2030 \cite{undrr2015sendai}, which identifies timely and inclusive early warning systems as a critical component of national resilience. Across many countries, mobile applications have emerged as essential components of the public warning ecosystem, complementing traditional channels such as broadcast media, SMS alerts, sirens, and agency websites. Disaster mobile apps offer several advantages: they can deliver geographically targeted warnings, integrate real-time hazard maps, support situational awareness, and provide preparedness advice before an event. These platforms offer richer, location-specific, and interactive communication than traditional channels. However, the effectiveness of these apps cannot be taken for granted. Their usefulness depends on aspects such as whether warnings are delivered on time and whether interfaces and disaster information are understandable. When alerts are delayed, maps are confusing, or interfaces are inaccessible, the consequences extend beyond poor usability; they can directly affect people’s ability to recognise danger and take protective action. These weaknesses also erode trust, potentially weakening protective action and increasing exposure to harm.

Despite their growing prominence in disaster risk communication, empirical understanding of the features offered by disaster warning apps and how the public experiences them during emergencies remains limited. Prior research has focused on developing new hazard applications or on usability evaluations of a small set of apps \cite{Tan2017, Navarro2021}. Research analysing user feedback from disaster warning applications is limited. Existing work relies on a relatively small number of user reviews (2,030 reviews), which restricts the generalisability of its findings \cite{tan2020usability}. As a result, emergency agencies and app developers lack comprehensive evidence on common design weaknesses, failure points, and communication barriers in disaster warning systems. To address this gap, this study conducts a large-scale analysis of 70 disaster mobile apps and 48,621 user reviews from the Google Play Store and Apple App Store. We examine the functional capabilities of these apps, identify user-reported challenges, and analyse public perceptions using topic modelling and sentiment analysis. This combination of feature analysis and large-scale user feedback provides valuable insights into how disaster mobile apps support or fail to support public preparedness and response during disasters. In addressing this, we have made the following contributions via this paper. 

\begin{itemize}
    \item Identify the disaster-related functions and characteristics of the apps, such as types of disasters covered, geographical areas covered, and monetisation strategies.
    \item Analyse user experiences and identify recurring challenges that affect the effectiveness of risk communication.
    \item Develop recommendations to make disaster mobile apps for inclusive and reliable.
\end{itemize}

By exploring the usability of these disaster mobile apps, this work contributes to ongoing discussions, such as the Early Warning For All by 2027 initiative \cite{wmo2022earlywarnings}, which aims to improve inclusiveness in digital disaster technologies. Our findings offer recommendations for emergency management agencies, policymakers, and developers seeking to build more resilient and accessible digital disaster communication systems.

\section{Background}

\subsection{Disaster Risk Communication and Early Warning Systems}
Effective disaster risk communication is crucial in enabling communities to anticipate disasters and interpret warnings, thereby allowing them to take timely protective action. As the natural hazards become more frequent, the pressure on emergency management agencies to deliver timely and reliable information has grown significantly. The Sendai Framework for Disaster Risk Reduction emphasises this need directly: Priority 1 calls for “understanding disaster risk,” while Priority 4 stresses the importance of “enhancing disaster preparedness for effective response,” including the provision of reliable, people-centred early warning systems \cite{undrr2015sendai}.

Disaster risk communication is more than sending alerts. It involves how quickly people receive it, whether they understand its meaning, and whether they trust the issuing authority enough to act upon it. Poorly communicated warnings, such as those that are delayed or confusing, can have serious consequences. Studies on major disaster events have shown that unclear messages reduce compliance and increase the likelihood of communities making life-threatening decisions during emergencies. In contrast, warnings that are timely and understandable are strongly associated with increased preparedness and faster protective action. Recent real-world incidents further illustrate these risks. During the 2025 California wildfires, an evacuation alert intended for a limited area was mistakenly sent to millions of residents across Los Angeles County, causing widespread confusion and anxiety before being retracted. Such alert failures highlight how technical errors and unclear targeting can undermine public trust and complicate emergency response efforts \cite{time_2025_la_evacuationalert, apnews_2025_la_alert_glitch}.

Modern early warning systems extend beyond traditional channels such as broadcast media, landline messages, sirens, and SMS alerts. With the widespread use of smartphones, mobile applications have become central to the warning ecosystem. These apps offer more than alerts; they support two-way information exchange, provide live maps, localised warnings, preparedness guidance, and real-time updates that help people understand what is happening around them. This positions mobile applications as a crucial platform for enhancing public situational awareness, particularly for rapidly evolving hazards such as bushfires, flash floods, cyclones, and severe storms.

However, the effectiveness of mobile-based warning systems depends strongly on their reliability and usability during high-stress situations. If alerts fail to trigger, if maps are unclear or inaccurate, or if interfaces become overwhelming during a crisis, users may miss critical information or misinterpret risk. Given this central role, there is a growing need to better understand how disaster mobile apps function in practice and how people experience them during real events.

\subsection{Mobile Applications in Disaster Preparedness, Response and Recovery}
Disaster mobile apps can be broadly classified into two categories: general-purpose apps and purpose-built apps. General-purpose apps include three subtypes: one-to-one, one-to-many, and many-to-many platforms \cite{Tan2017}. One-to-one channels, such as WhatsApp and WeChat, support private communication between individuals. One-to-many platforms include news apps (e.g., CNN, BBC), which mainly report general news but also share hazard updates. Many-to-many platforms such as Twitter and Facebook enable wide community sharing during emergencies. Research shows that people often rely on these familiar platforms during crises because they are already embedded in everyday communication practices \cite{HADDOW201453}.

Despite their wide use, emergency agencies remain cautious about depending on general-purpose platforms for official disaster communication. Concerns include privacy, misinformation, content quality, and the lack of control over message consistency \cite{Tan2017}. For this reason, many countries invest in purpose-built disaster mobile apps, which are tools developed and maintained by emergency management authorities. These apps focus on issuing authorised warnings, providing verified updates, and supporting situational awareness through structured, reliable information flows. They are typically one-way systems, where information is pushed from authorities to the public. Examples include Germany’s NINA app and the FEMA app in the United States, both of which are designed to provide multi-hazard alerts and official guidance.

While purpose-built apps offer greater control, they also face challenges. Their effectiveness depends on the timely ingestion of data and clear communication design. Users often compare their experience with purpose-built apps to general-purpose platforms, particularly social media, which can feel more responsive and interactive. Understanding how these apps operate and how people experience them during disasters is crucial for enhancing the reliability and inclusivity of mobile-based disaster communication systems.

\subsection{Usability, Reliability, and Inclusiveness Challenges in Disaster Mobile Apps}
Although disaster mobile apps offer clear benefits, their effectiveness depends heavily on how well they perform during high-pressure situations. Disasters place unique cognitive and emotional demands on users. People often interact with apps while stressed, time-poor, or when they are navigating rapidly changing conditions. As a result, even minor usability or reliability issues can significantly impact how people perceive risk and whether they take protective action. 

Studies consistently demonstrate that \textit{usability} is crucial to effective risk communication. Interfaces with cluttered layouts, low-contrast text, ambiguous icons, or complex navigation increase cognitive load, making it more difficult for users to quickly understand hazard information. The Modified Usability Framework for Disaster Mobile Apps highlights that intuitive design, minimal cognitive effort, and clear visual hierarchy are essential for supporting protective behaviour under stress \cite{tan2020understanding}. 
\textit{Reliability} is another equally important aspect. Delayed alerts, inconsistent notifications, missing data, app crashes, and slow loading times directly affect a user’s ability to recognise danger. Technical failures during emergencies are not just an inconvenience; they compromise situational awareness and place a risk on lives.
\textit{Inclusiveness} is another significant challenge. Disaster mobile apps often assume a high level of digital literacy, English proficiency, or visual ability. Many apps lack support for older adults, users with disabilities, or culturally and linguistically diverse communities \cite{madugalla2025human}. Issues such as small text, inaccessible colour schemes, poor screen-reader support, limited language options, and complex navigation disproportionately affect already vulnerable groups \cite{Designin57_online}. Past disasters show that when warning systems fail to meet the needs of diverse users, inequalities in preparedness and response widen \cite{budimir2020nepal, amaratunga2022genderEWS, perera2020societal}. In summary, these issues highlight that disaster mobile apps performance is not only a technical matter but a critical component of risk communication and public safety. Understanding the specific user challenges as reported by users provides important insights for improving future disaster mobile apps design and strengthening early warning systems more broadly.

\subsection{User Feedback App Reviews as a Source of Insight}
Understanding how disaster mobile apps perform in real-world conditions requires insights from the people who rely on them during emergencies. While technical evaluations and controlled usability studies provide valuable information, they often cannot capture the full range of challenges that emerge during actual hazard events. In contrast, user-generated reviews on app stores such as the Google Play Store and Apple App Store have been shown to provide unique insight into how apps perform in areas such as medical \cite{perera2022mhealth}, banking \cite{amirkhalili2025bankingonfeedback}, and travel \cite{caylak2024travel}.

App reviews are written by users who interact with these apps in diverse contexts, including high-stress situations such as during disasters. These reviews often describe issues that are difficult to observe in laboratory settings, such as slow updates, crashes, battery drain, or issues in understanding information. Reviews also reveal emotional responses such as frustration or a loss of trust, which influence how people interpret warnings and respond to risk \cite{Tan2020Modified, khans2024softwareInsights}. Reviews further provide insight into recurring patterns of problems across different apps. When many users report similar failures—such as delayed notifications or unclear maps, this signals broader systemic weaknesses rather than isolated defects. In the disaster context, such information will be particularly valuable for emergency agencies seeking to understand how their apps compare with user expectations and with the responsiveness of general-purpose platforms, such as social media. User feedback is also a rich source of information about inclusiveness and accessibility. Reviews frequently describe difficulties related to language, visual design, readability or cognitive load. Such issues disproportionately affect older adults, people with disabilities, and culturally and linguistically diverse communities. App reviews, therefore, complement traditional accessibility evaluations by providing direct accounts from users with varied needs and levels of digital literacy. Due to these reasons, exploring app reviews in the context of disaster mobile apps will provide useful insights for future disaster mobile apps development projects. 

Beyond the DRR domain, app reviews have also received substantial attention in software engineering research. Given that reviews contain rich descriptions of real user experiences, many studies have mined them to extract actionable insights for requirements engineering and quality improvement tasks \cite{dkabrowski2022analysing,genc2017systematic}. Researchers have used review mining to identify feature requests \cite{johann2017safe,de2021re,Motger2024}, extract non-functional requirements such as performance and usability concerns \cite{jha2019mining,lu2017automatic}, and determine which features users value most or critique most strongly \cite{Wu2021}. Additional studies have focused on detecting issues such as UI problems, crashes, and bugs \cite{man2016experience,chen2021should,maalej2015bug,mcilroy2016analyzing}, or identifying fake and low-quality reviews \cite{martens2019towards,noei2019too}. Sentiment analysis has been widely applied to capture users’ emotional responses to apps \cite{martens2017emotion,luiz2018feature}, and more recent work has examined human, social, and ethical concerns embedded in reviews \cite{shahin2023study,obie2023automated,shams2020society,bowie2022user,nema2022analyzing}.

This body of work demonstrates that user reviews offer a credible and valuable source of data for understanding how disaster mobile apps function in practice. When analysed at scale, they can reveal widespread pain points, communication gaps, and opportunities for improvement. Leveraging this type of user-generated evidence helps build a more complete picture of disaster mobile apps performance and supports the development of more user-centred disaster communication tools.

\subsection{Text Mining and Sentiment Analysis in Disaster Research}
Text mining has become an important approach in disaster research, enabling the analysis of large volumes of unstructured data generated by the public during emergencies. Social media posts, online comments, and user reviews contain first-hand accounts of how people perceive hazards, interpret warnings, and experience emergency communication systems. Analysing this text at scale allows researchers to identify emerging concerns, track public sentiment, and detect communication gaps that may influence protective action.

Several studies have demonstrated the value of these methods for understanding disaster-related behaviour. For example, Roy et al. \cite{roy2020socialmedia} analyze 52.5 million tweets related to Hurricane Sandy to assess the effectiveness of social media communication during disasters and identify the contributing factors leading to effective crisis communication strategies. Similar studies on extreme weather events have used topic modelling, clustering, and temporal analysis to explore how people discuss risk and how public attention shifts as situations evolve \cite{riskPerception2025, dahal2019topic, chen2023tracking, florenceSpatiotemporal2020}.

In the context of disaster mobile apps, text mining provides a powerful means of uncovering user-reported challenges and expectations that may not be captured through technical evaluations. Topic modelling can identify recurring themes, such as alert delays, map issues, or trust concerns, by automatically grouping similar reviews. This approach is particularly useful when analysing content from multiple countries, apps, and hazard types, where the reviews are numerous and the patterns are diverse and complex. Sentiment analysis complements this by estimating the emotional tone associated with different themes, providing insight into which issues generate frustration, confusion, or distress \cite{ramzy2024covidAppSentiment, li2020sentimentStatistical}. These emotional responses are relevant to disaster communication, as negative experiences can reduce trust in official apps and discourage continued use. For disaster mobile apps, sentiment analysis can highlight features or failures perceived as most harmful, such as unreliable warnings or unclear hazard boundaries. 

Together, topic modelling and sentiment analysis offer a systematic way to examine large-scale user experiences and translate these insights into evidence-based improvements. However, as this has not been conducted on disaster mobile apps reviews before, this is the first study that seeks to identify user challenges that influence the effectiveness of disaster mobile apps. These findings form the basis for the design recommendations presented later in this paper.

\section{Methodology}
\begin{figure*}
	\centering
		\includegraphics[width=\textwidth]{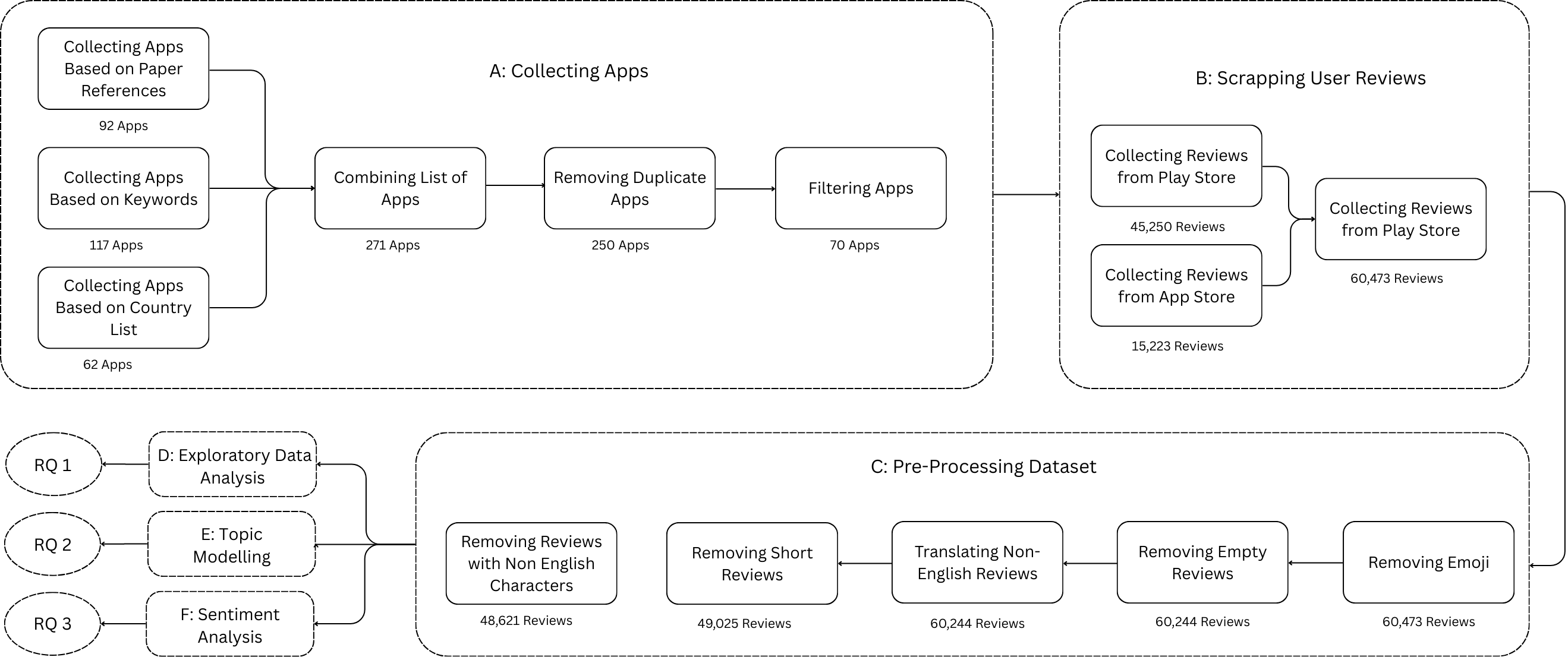}
	\caption{Overview of our study approach}
	\label{FIG_overview_of_study}
\end{figure*}

This study adopts a mixed‐method computational approach combining (1) a feature analysis of disaster mobile apps and (2) large-scale text mining of user reviews. The aim is to understand how these apps support disaster preparedness and response, what challenges users report during real events, and how these insights can inform the design of more reliable and inclusive disaster communication tools. We structured the methodology around three research questions. To address these questions, we followed the multi-stage process outlined in Figure \ref{FIG_overview_of_study} and described in the next few sections. The research questions and their rationale for selection are provided below. \\

\noindent \textbf{RQ1: How is the Disaster Mobile Apps Landscape structured?}

\noindent \textit{Rationale: }This question examines the fundamental properties of disaster mobile apps, including the features they provide, the types of natural hazards they address, and the geographic regions they cover. We also analyse indicators of app uptake and perceived quality, such as download counts and user ratings. Where relevant, we consider monetisation strategies to understand how these apps are positioned in the public domain. Addressing RQ1 helps emergency agencies and developers identify which features are currently offered and which features should be prioritised in future disaster mobile apps design. \\

\noindent \textbf{RQ2: What do users discuss in disaster mobile apps reviews?}

\noindent \textit{Rationale: } User reviews contain diverse perspectives on how disaster mobile apps function during real events. To capture these perspectives, we apply topic modelling to identify and categorise the main themes discussed in 48,621 reviews across 70 apps. This allows us to uncover the issues that matter most to users, including both functional and communication-related concerns. Understanding these themes provides insight into user expectations and the challenges they encounter in practice. \\

\noindent \textbf{RQ3: What are the common challenges in using disaster mobile apps as reported by their users?}

\noindent \textit{Rationale: } This question focuses specifically on user-reported problems that may influence trust, usability, and the effectiveness of disaster communication. We use a combination of user ratings and sentiment analysis to assess the emotional tone of these issues. This enables us to identify which challenges cause the greatest frustration or distress and therefore pose the highest risks to public safety and decision-making.

\subsection{App Identification and Selection}
To build a comprehensive dataset of disaster mobile apps, we conducted a systematic, multi-step collection process in October 2023, and it was later refreshed in January 2026 to include the latest apps. Applications published after this were not included. The process involved finding apps using various techniques and filtering them to produce a refined set.

\subsubsection{Collecting Apps}
We used three complementary approaches designed to capture both well-known and lesser-known apps across different countries. This three-stage process helped to ensure that a large collection of apps was gathered. \\
\noindent \textit {\textbf{Approach 1: Literature-based app collection: }}We reviewed existing academic studies that compiled disaster management applications. From two key papers \cite{Tan2017, Tan2020Modified}, we extracted a combined list of 92 mobile apps. \\
\noindent \textit {\textbf{Approach 2: Keyword-based app collection: }}To expand beyond literature coverage, we searched the Google Play Store and Apple App Store using disaster-specific keywords such as ‘flood’, ‘earthquake’, and ‘tornado’ as per Navarro et. al.'s study \cite{Navarro2021}. These keywords were used to search in the titles and descriptions of apps in the Google Play Store and Apple App Store. This process identified 117 additional apps.\\
\noindent \textit{\textbf{Approach 3: Country-based app collection: }}Keyword searches did not capture all disaster mobile apps—especially those whose names or descriptions do not explicitly reference hazards. For instance, ‘VicEmergency’, an app developed by an Australian state government, provides a variety of warnings, including those for natural disasters. However, its name and tags lack any direct reference to ‘disaster’ or specific disaster types. The description of `VicEmergency' merely states it as the official app for emergency warnings and information in Victoria, with tags limited to ‘News and Emergency’. Additionally, apps developed in non-English speaking countries often have names in their native languages, making them hard to find through English keyword searches, such as the German government's app `NINA - Die Warn-App des BBK'. To address this, we conducted a country-by-country Google search using the query ``Emergency apps in [Country]”. This query was chosen because emergency apps often include hazard-related features even when labelled differently. Using this method, we identified additional apps, such as Japan’s \textit{Yurekuru Call and Safety Tips}, which were located through a blog by Robin Lewis \cite{lewis:2018}. We repeated this process for all 196 countries listed by Britannica \cite{britannica}. By following approach 3, we identified 62 apps. We were hoping this number to be higher to reflect the fact that there are 196 countries in the world. This relatively small number is due to the fact that many countries do not rely on dedicated mobile applications for warning dissemination, but instead use alternative communication channels. For instance, Austria utilises sirens \cite{austria_warning_bmi}, Canada informs their citizens via television and radio \cite{alertready_canada}, India employs SMS \cite{ndma_sachet_2025}, and Belgium uses email \cite{bealert_belgium_2025}. These examples represent only one of the methods used in each country, as most nations adopt multi-channel warning systems that combine several communication mechanisms to disseminate disaster information.

With these three approaches, we found 271 disaster mobile apps. However, there were several duplicates, and after merging these lists and removing duplicates (see Figure \ref{FIG_overview_of_study}), the resulting list contained 250 apps. 

\subsubsection{Filtering Apps} 
For the 250 apps, we applied two filtering processes to refine this collection. The first was an \textit{availability filter}. This meant that apps that were unavailable in the Google Play Store or Apple App Store were removed, as their reviews were inaccessible. As of January 2026, 72 apps were unavailable on the stores, leaving us with 178 apps. The second filter was a \textit{false positive filter}. False positives were defined as applications that did not provide functionalities directly related to natural disaster management, such as disaster alerts, real-time disaster maps, or disaster preparedness guidance. Examples of excluded applications include those focused on crime-related emergencies (e.g., robbery) and medical emergencies (e.g., cardiac arrest). Although such applications address emergency situations, they fall outside the scope of this study, which focuses exclusively on natural disasters. This false-positive criterion also applies to generic social media and crowdfunding platforms, which may contain incidental disaster-related information but are not explicitly designed for disaster management purposes. To ensure the accuracy of this filtering stage, we downloaded and manually inspected a subset of the applications to verify their functionality. After applying this filter, 70 apps remained. The complete list of selected applications is presented in Appendix Table~\ref{tab:apps}.

\subsection{Understanding the Disaster Mobile Apps Landscape}
After constructing the final dataset of 70 disaster mobile apps, we conducted an exploratory analysis to address \textit{RQ1: How is the disaster mobile apps landscape structured?}. This analysis aimed to provide an overview of the characteristics of existing disaster mobile apps and to identify common patterns in their design and deployment. Basic application metadata, including pricing schemes, number of downloads, and user ratings, was collected directly from the official app pages on the Google Play Store and Apple App Store. To gain deeper insights into each application's functional capabilities, we also conducted a hands-on evaluation by downloading and interacting with each app. This process allowed us to examine aspects that are not always visible in app store descriptions or screenshots, such as supported disaster types, geographic coverage, and operational features. Based on this analysis, we classified each application along five key dimensions.

\textbf{D1: Type of Disasters (Disaster Scope)}  
Apps were first categorised according to the types of disasters they support. Applications focusing on a single hazard (e.g., earthquakes only) were classified as \textit{single-disaster} apps, whereas applications providing alerts or information for multiple hazards (e.g., floods, bushfires, storms, and heatwaves) were classified as \textit{multi-disaster} apps.

\textbf{D2: Geographic Coverage}  
    We then analysed the geographic regions supported by each application using information provided in the app descriptions and configuration settings. Apps were categorised into four coverage types:
    \begin{itemize}
        \item Local: Coverage limited to sub-national regions such as states, provinces, territories, or cities.
        \item National: Applications designed to serve users within a single country.
        \item Regional: Applications covering disaster events across multiple countries or geographic regions (e.g., ocean basins).
        \item Global: Applications that provide worldwide coverage and are usable across multiple countries.
    \end{itemize}
    
This classification helps reveal how disaster information services are distributed geographically and whether applications are designed for localised or international disaster monitoring.

\textbf{D3: Popularity and Perceived Quality}  
We used download counts and user ratings as indicators of application adoption and perceived quality. Download statistics provide insight into how widely an app is used, while user ratings offer a coarse indicator of user satisfaction. These values were obtained directly from the respective app store listings.

\textbf{D4: Monetisation Strategy}  
We analysed the monetisation approach adopted by each application using information available on the app store pages. Apps were categorised into the following pricing models: \textit{Free}, \textit{Free with advertisements}, \textit{Free with in-app purchases}, \textit{Free with advertisements and in-app purchases}, \textit{Paid}, and \textit{Paid with in-app purchases}. Understanding monetisation strategies is important because advertisements, subscription models, or paywalls may affect usability and accessibility, particularly in time-critical disaster situations.

\textbf{D5: Feature Analysis by Disaster Lifecycle Stages}
To understand the functional capabilities of disaster mobile apps, we conducted a hands-on feature analysis by interacting directly with each application. During this process, we documented all disaster-related functionalities observed during real use of the app interface. The identified features were then categorised according to the stages of the disaster management lifecycle: \textit{preparation}, \textit{response}, \textit{recovery}, and \textit{mitigation}. This categorisation allows us to examine how current applications support different phases of disaster management and to identify stages that may be underrepresented in existing solutions.

These five dimensions collectively provide a structured view of the current disaster mobile apps landscape. The results of this analysis are presented in Section~\ref{sec:results-landscape} and form the basis for interpreting user feedback in later sections.

\subsection{User Review Collection and Pre-processing}
To address RQ2 and RQ3, we collected user reviews associated with the selected disaster mobile apps and performed several preprocessing steps to prepare the dataset for analysis.

\textbf{Review Collection:}
User reviews for the 70 selected applications were collected from both the Google Play Store and the Apple App Store. Data collection was conducted using two publicly available scraping tools: an app-store-scraper \footnote{\url{https://github.com/cowboy-bebug/app-store-scraper}} and google-play-scraper \footnote{\url{https://github.com/JoMingyu/google-play-scraper}}.  Using these tools, we retrieved all publicly available user reviews associated with each application. In total, we collected 45,250 reviews from the Google Play Store and 15,223 from the Apple App Store, for a combined dataset of 60,473 reviews.

\textbf{Emoji Removal and Empty Review Filtering:}
The initial dataset contained reviews that consisted entirely of emojis or other non-textual characters. Because topic modelling relies on semantic textual content, emojis were removed during preprocessing. After removing emojis, 229 reviews became empty and were therefore excluded from the dataset. The remaining reviews, a total of 60,244 reviews, were retained for further processing.

\textbf{Language Detection and Translation:}
The collected reviews were written in multiple languages, particularly for applications developed in non-English-speaking countries. For example, applications such as \textit{Alertswiss} (Switzerland), \textit{Emergency Ready} (South Korea), and \textit{Yurekuru Call} (Japan) contained a substantial number of reviews in local languages. In addition, global applications also contained a substantial number of user reviews written in languages other than English. 

To ensure consistent analysis, we first performed automatic language detection using the \textit{langdetect} library \cite{langdetect}. This process identified 7,947 reviews written in languages other than English. These reviews were then translated into English using the OPUS-MT zero-shot translation model \cite{Tiedemann2023}. OPUS-MT was selected due to its extensive multilingual coverage and strong translation performance. The OPUS-MT repository contains more than 2,300 machine translation models supporting 4,560 translation directions across 294 languages \cite{Tiedemann2023}. Previous studies have also shown that its translation performance is competitive with large multilingual models such as NLLB \cite{nllbteam2022languageleftbehindscaling}.

\textbf{Removing Short Reviews and Reviews with Non-English Characters:}
Very short reviews often contain limited semantic information (e.g., ``good'', ``not accurate'', ``works well'') and therefore contribute little to topic modelling. To improve topic quality, reviews consisting of three words or fewer were removed. After applying this filter, the dataset was reduced to 49,025 reviews. 

Finally, reviews containing non-Latin characters (e.g., Bengali or Arabic scripts) were removed. Although translation was applied, language-detection tools may not reliably identify the source language for certain scripts, potentially introducing noise into downstream analysis. Excluding these entries helped ensure the reliability of the topic modelling and sentiment analysis stages.

\subsection{Topic Modelling}

\begin{figure}
    \centering
    \includegraphics[width=0.8\textwidth]{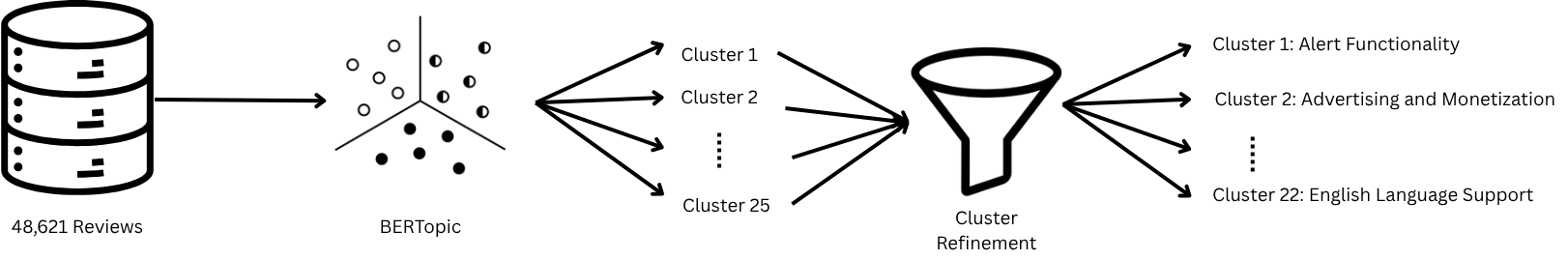}
    \caption{Topic Modeling Process}
    \label{fig:topic_modelling_process}
\end{figure}

To address RQ2, we applied topic modelling to identify the main themes discussed in user reviews of disaster mobile apps. Topic modelling enables the discovery of latent semantic patterns in a large text corpus without requiring manual labelling. The overall topic modelling workflow is illustrated in Figure \ref{fig:topic_modelling_process}.

We used BERTopic \cite{grootendorst2022bertopic}, a topic modelling framework that combines transformer-based language models with density-based clustering techniques to generate semantically coherent topics. First, each review was converted into a vector representation using Sentence-BERT (SBERT) embeddings \cite{reimers2019sentencebertsentenceembeddingsusing}. These embeddings capture contextual semantic relationships between sentences, allowing reviews with similar meanings to be positioned closer together in vector space. Because the resulting embeddings are high-dimensional, dimensionality reduction was performed using Uniform Manifold Approximation and Projection (UMAP) \cite{mcinnes2020umap}. UMAP preserves both local and global structure in the data while reducing computational complexity for clustering. The reduced embeddings were then clustered using Hierarchical Density-Based Spatial Clustering of Applications with Noise (HDBSCAN) \cite{mcinnes2017hdbscan}. HDBSCAN groups semantically similar reviews into clusters while automatically identifying noise points that do not strongly belong to any topic. To determine an appropriate clustering configuration, we evaluated clustering quality using the silhouette coefficient \cite{Rousseeuw1987}, where higher values indicate better separation between clusters.

\subsection{Sentiment Analysis}
\label{subsec:sentiment_analysis}
To address RQ3, we analysed the sentiment associated with user discussions in disaster mobile apps reviews. This analysis aimed to identify the areas where users experience the most dissatisfaction, thereby highlighting areas that require urgent attention from developers and emergency agencies.

Arguably, user ratings can also be used to understand user sentiment/satisfaction with an app. However, relying solely on ratings may lead to inaccurate interpretations due to the phenomenon known as \textit{Text–Rating Inconsistency} (TRI) \cite{BinFu2013}. TRI can occur when users assign ratings that do not reflect the sentiment expressed in the accompanying review text \cite{BinFu2013}. For example, a user may provide a high rating immediately after installing an app without meaningful use, whereas another user may express multiple complaints but still assign a neutral rating, such as three stars. Relying solely on numerical ratings can therefore obscure real user experiences and introduce noise into supervised models.

\begin{figure}
	\centering
		\includegraphics[width=0.6\textwidth]{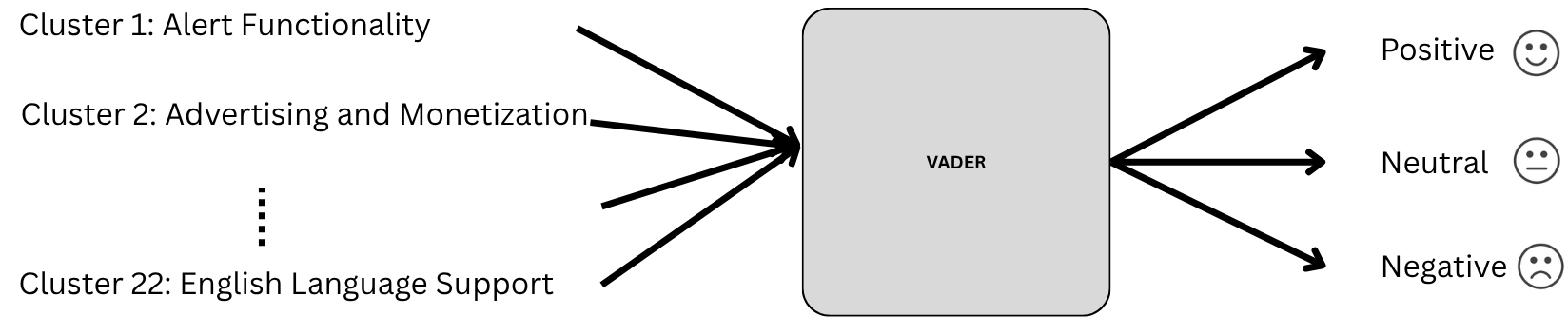}
	\caption{Sentiment Analysis Process}
	\label{FIG_sentiment_analysis_process}
\end{figure}

To avoid this issue, we adopted an unsupervised sentiment analysis approach that does not rely on the accuracy of user rating labels. We use VADER \cite{hutto2014vader}, a sentiment analysis tool designed for short, informal user-generated text. The overall process followed in this study is shown in Figure~\ref{FIG_sentiment_analysis_process}. VADER produces four sentiment scores for each review: \textit{positive}, \textit{neutral}, \textit{negative}, and a \textit{compound} score that represents the overall sentiment of the text. Because a review may contain both positive and negative elements, the compound score aggregates these signals into a single value ranging from $-1$ (most negative) to $+1$ (most positive). For instance, the review \textit{``Currently, I like the app very much, but you should be able to change the language within the app.''} contains both positive (``I like the app very much'') and negative (``you should be able to change the language'') components. VADER captures this by assigning both positive and negative values, and then computing a compound score ranging from $-1$ to $+1$.

\textbf{Distribution of Review Sentiment: } We first analysed the overall distribution of sentiment across all reviews. This provided an overview of how users emotionally responded to disaster mobile apps. VADER generates a compound score ranging from -1 to 1. To convert this score into three sentiment classes, we classify reviews with compound scores into three classes, Positive: compound score $\geq 0.33$, Neutral: $-0.33 < $ compound score $ < 0.33$, Negative: compound score $\leq -0.33$.

\textbf{Topic-Level Sentiment: } Next, we analysed sentiment at the topic level by combining the sentiment scores with the topics identified through BERTopic. For each topic, we computed the average sentiment score and the average user rating. This analysis allowed us to identify which topics are associated with the most negative user experiences, enabling developers and emergency agencies to prioritise improvements in those areas.

In summary, sentiment analysis and topic modelling provide a more reliable understanding of user experience than ratings alone, helping surface the challenges that most affect usability and the effectiveness of disaster mobile apps.

\section{Results}
This section presents the findings of our study based on the analysis of 70 disaster mobile apps and 48,621 user reviews. The results are organised according to the research questions introduced in Section 3. First, we analysed the characteristics of the disaster mobile apps landscape (RQ1). We then examined the topics discussed in user reviews (RQ2) and finally analysed the sentiment associated with these topics to identify key user challenges (RQ3).

\subsection{Understanding the Disaster Mobile Apps Landscape (RQ1)}
\label{sec:results-landscape}

\subsubsection{Disaster Scope}
Disaster mobile apps differ in the range of hazards they address. Based on the types of disasters supported, we classified apps into two categories: \textit{single-disaster} and \textit{multi-disaster} applications. Single-disaster mobile apps focus on monitoring or issuing alerts for a single hazard type. In our dataset, these apps mainly focused on earthquakes, wind-related disasters (e.g., hurricanes and cyclones), floods, or fires. For example, the \textit{LastQuake} application provides alerts and information specifically about earthquakes. In contrast, multi-disaster mobile apps support multiple hazard types within a single platform. For instance, the \textit{Disaster Alert} app provides information on earthquakes, floods, storms, and other hazard events. Table \ref{tab_disaster_type} shows the distribution of apps by disaster scope.
\begin{table*}[t]
\caption{\small Number of Apps by Disaster Type}
{\label{tab_disaster_type}
\centering
\begin{tabular}{@{}ll@{}}
\toprule
\textbf{Type of Disaster} & \textbf{Number of Apps} \\ \midrule
Multi-Disasters & \mybarhhigh{26}{26} \\
\midrule
Earthquakes & \mybarhhigh{25}{25} \\
Wind Disaster & \mybarhhigh{12}{12} \\
Floods & \mybarhhigh{6}{6} \\
Fire Disaster & \mybarhhigh{1}{1} \\
\bottomrule
\end{tabular}
}
\end{table*}

\subsubsection{Geographic Coverage}
Disaster mobile apps vary significantly in the geographic areas they serve. Based on our analysis, we classified apps into four geographic coverage categories: \textit{local}, \textit{national}, \textit{regional} and \textit{global}. Local applications target smaller geographic areas such as states or cities; for instance, \textit{VicEmergency} focuses on the state of Victoria in Australia. National applications focus on a single country, such as \textit{Yurekuru Call}, which provides earthquake alerts specifically for Japan. Regional applications monitor disasters occurring across several ocean basins, such as \textit{Tropical Hurricane Tracker}, which monitors the Atlantic, Central Pacific, and Eastern Pacific regions. Global applications provide disaster monitoring for multiple countries distributed across the globe. For example, the \textit{Disaster Alert} app provides hazard information for many regions worldwide. Table \ref{tab:geo_coverage} presents the distribution of apps across these categories. The results show that global applications form the largest group, followed by national apps. 

\begin{table}[H]
\caption{\small Number of Apps by Geographic Coverage}
{\color{black}
\label{tab:geo_coverage}
\centering
\begin{tabular}{@{}ll@{}}
\toprule
\textbf{Geographic Coverage} & \textbf{Number of Apps} \\ \midrule
Global & \mybarhhigh{34}{34} \\
Regional & \mybarhhigh{4}{4} \\
National & \mybarhhigh{21}{21} \\
Local & \mybarhhigh{11}{11} \\
\bottomrule
\end{tabular}
}
\end{table}

\subsubsection{Popularity and Perceived Quality}
The popularity and perceived quality of disaster mobile apps were assessed using download counts and user ratings on the Apple App Store and Google Play Store. Table \ref{tab:downloads_ratings_summary} summarises the descriptive statistics for both platforms. Regarding \textit{app popularity}, download counts vary considerably between platforms. On the Apple App Store, the mean number of downloads was approximately 16K, with a median of 2K. Downloads ranged from a minimum of 10 to a maximum of 111K. In comparison, the Google Play Store showed higher adoption, with a mean of 700K downloads and a median of 100K, ranging from 5K to 10M downloads. 

User ratings, reflecting \textit{perceived quality}, were generally high on both platforms. The Apple App Store had a mean rating of 4.07 and a median of 4.6, with a minimum of 1.5 and a maximum of 4.9. Ratings on the Google Play Store are slightly lower, with a mean of 3.91 and a median of 4.2, ranging from 2.1 to 4.9. These statistics indicated that most disaster mobile apps were positively perceived by users, despite differences in adoption between platforms.

\begin{table}[H]
\centering
\caption{Descriptive Statistics of Downloads and Ratings for Apple App Store and Google Play Store}
{\color{black}
\label{tab:downloads_ratings_summary}
\begin{tabular}{llcc}
\toprule
\textbf{Variable} & \textbf{Statistic} & \textbf{Apple App Store} & \textbf{Google Play Store} \\
\midrule
Download & Mean & 16K+ & 700K+ \\
         & Median & 2K+ & 100K+ \\
         & Min & 10 & 5K+ \\
         & Max & 111K+ & 10M+ \\
\midrule
Rating   & Mean & 4.07 & 3.91 \\
         & Median & 4.6 & 4.2 \\
         & Min & 1.5 & 2.1 \\
         & Max & 4.9 & 4.9 \\
\bottomrule
\end{tabular}
}
\end{table}

\subsubsection{Monetisation Strategy}

We analysed the monetisation strategies of disaster mobile apps on the Apple App Store and Google Play Store. Table~\ref{tab:monetisation} summarises the distribution of pricing schemes across both platforms. Most disaster mobile apps were offered free of charge. On the Apple App Store, most apps were either fully free (19) or used a freemium model with in-app purchases (16). Only a small number of apps used a paid model with in-app purchases (2), while no apps relied solely on paid access or advertising. In contrast, the Google Play Store offered a wider variety of monetisation approaches, including free apps supported by ads and apps that combine ads with in-app purchases. Most apps were free (29), but several included advertisements (8), in-app purchases (3), or both (13). Paid apps were also present, either as standalone paid apps (3) or as paid apps with in-app purchases (2).

These results indicate that free access is the predominant monetisation strategy on both platforms. The Google Play Store showed greater use of ads and combined monetisation approaches, suggesting a higher prevalence of revenue-generation strategies integrated with free access. In comparison, Apple App Store apps relied more on the freemium model through in-app purchases. The presence of advertisements and subscription models can raise potential concerns regarding usability and accessibility during time-critical disaster situations.
 
\begin{table}[H]
\centering
\caption{Monetisation Strategies of Disaster Apps on Apple App Store and Google Play Store}
{\color{black}
\label{tab:monetisation}
\begin{tabular}{lcc}
\toprule
\textbf{Pricing Scheme} & \textbf{Apple App Store} & \textbf{Google Play Store} \\
\midrule
Free & 19 & 29 \\
Free with Ads & 0 & 8 \\
Free with In-app Purchase & 16 & 3 \\
Free with Ads and In-app Purchase & 0 & 13 \\
Paid & 0 & 3 \\
Paid with In-app Purchase & 2 & 2 \\
\bottomrule
\end{tabular}
}
\end{table}

\subsubsection{Feature Analysis by Disaster Lifecycle Stages}
To understand how disaster mobile apps supported the different phases of disaster management, we analysed the features offered by each application and mapped them to the disaster management lifecycle stages: preparation, response, and recovery \cite{ha2026reviewing}. The number of apps offering each feature type is shown in Table \ref{tab:feature_distribution}. The results show that most applications focused heavily on response-related features, such as real-time alerts and disaster maps. Preparation features, including forecasting tools and preparedness guidance, were less common, while recovery-related features were present only in a few apps. This distribution suggests that current disaster mobile apps prioritise real-time situational awareness during events, while providing less support for pre-disaster preparation and post-disaster recovery. In the next few sections, we provide more details on the features supported in each stage. 

\begin{table}[H]
\centering
\caption{Distribution of Features Across Disaster App Categories}
{\color{black}
\label{tab:feature_distribution}
\begin{tabular}{llc}
\toprule
\textbf{Category} & \textbf{Feature} & \textbf{Number of Apps} \\
\midrule
Preparation & Early Warning Alerts & 23 \\
            & Disaster Forecasting & 17 \\
            & Preparation Tips     & 12 \\
\midrule
Response    & Real Time Alerts     & 53 \\
            & Disaster Maps        & 57 \\
            & Real Time Tracking   & 11 \\
            & Emergency Reporting  & 7  \\
\midrule
Recovery    & List of Disaster     & 51 \\
            & News / Bulletin      & 13 \\
            & Share Experiences    & 13 \\
            & Safety Confirmation  & 5  \\
            & Find Shelter         & 5  \\
\bottomrule
\end{tabular}
}
\end{table}

\paragraph{\textbf{Preparation Stage: }} Features in this stage focused on helping users prepare before a disaster occurred. 

\begin{figure*}
	\centering
		\includegraphics[width=0.5\textwidth]{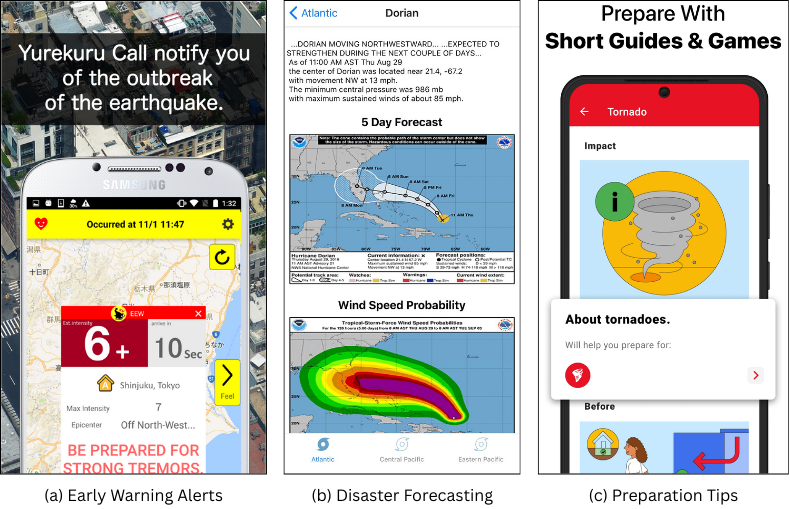}
	\caption{Features of the Preparation Stage}
	\label{preparation}
\end{figure*}

\begin{enumerate}
    \item \textbf{Early Warning Alerts}: These alerts notify users shortly before a disaster, often minutes or hours in advance, allowing them to take precautionary measures. This feature was particularly common in earthquake monitoring applications developed in Japan, such as the \textit{Yurekuru Call} app (Figure \ref{preparation}a).

    \item \textbf{Disaster Forecasting}: This feature supported the prediction of potential disasters at specific times and locations. Unlike early warning alerts, which typically occur shortly before an event, forecasting provides a longer lead time but with lower accuracy. This functionality was commonly found in applications monitoring wind-related disasters such as hurricanes and typhoons. An example is the \textit{Seastorm Hurricane Tracker} app (Figure \ref{preparation}b).

    \item \textbf{Preparation Tips/Tutorials}: These features provide educational and training material to help users prepare for disasters. It included self-rescue tips, tools needed, and guidance for post-disaster response. For instance, the \textit{Emergency: Severe Weather App} provided preparedness instructions for tornado events (Figure \ref{preparation}c).

\end{enumerate}

\paragraph{\textbf{Response Stage: }} Features in this stage supported users during an ongoing disaster

\begin{figure*}
	\centering
		\includegraphics[width=0.6\textwidth]{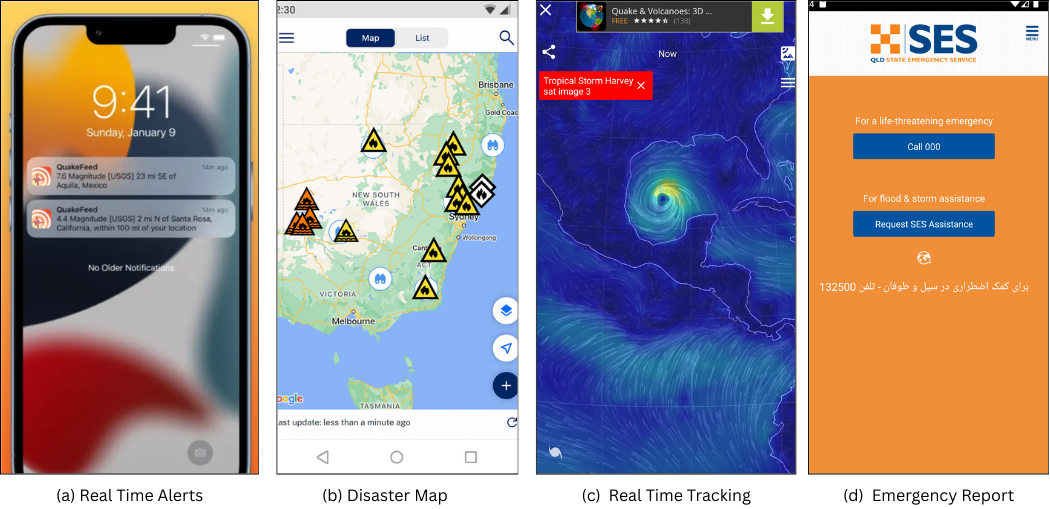}
	\caption{Features of the Response Stage}
	\label{response}
\end{figure*}

\begin{enumerate}
    \item \textbf{Real-Time Alerts}: With this feature, notifications will appear during the disaster, informing the user about events occurring in their area or within a specified radius. For example, if a user sets a threshold for earthquakes with a magnitude of 3 or higher on the Richter scale, they will receive a notification when such an event occurs. An example is the \textit{QuakeFeed Earthquake Tracker} app (see Figure \ref{response}a).
        
    \item \textbf{Disaster Maps}: This feature provided geographic visualisations typically through interactive maps displaying hazard locations and affected areas. An example is the \textit{Hazards Near Me NSW} app (Figure \ref{response}b).
    
    \item \textbf{Real-Time Tracking}: Some apps offered live updates on ongoing disasters or wind conditions. This was commonly used for wind-related disasters such as hurricanes, cyclones, and typhoons, often displaying animated representations of events. An example is the \textit{Wind Map Hurricane Tracker 3D} app (see Figure \ref{response}c).
    
    \item \textbf{Emergency Reporting}: Certain apps allowed users to contact emergency services or report incidents directly through the application interface. For example, the \textit{SES Assistance QLD} app provided a feature for requesting emergency assistance (Figure \ref{response}d).

\end{enumerate}

\paragraph{\textbf{Recovery Stage:}} Features in this stage supported users after a disaster event.

\begin{figure*}
	\centering
		\includegraphics[width=0.8\textwidth]{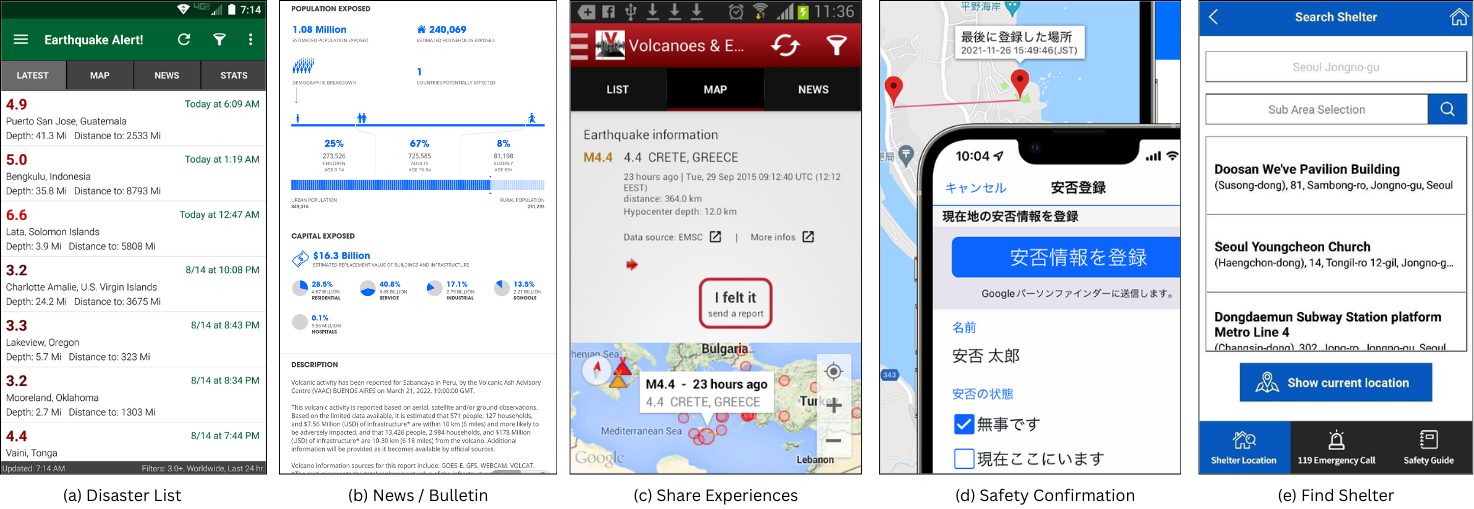}
	\caption{Features of the Recovery Stage}
	\label{recovery}
\end{figure*}

\begin{enumerate}
    \item \textbf{Disaster List}: Presented concise updates, often in list format, showing statistics such as magnitude, depth, and proximity to cities. An example is the \textit{Earthquake Alert!} app (see Figure \ref{recovery}a).
    
    \item \textbf{News/Bulletins}: Provided comprehensive post-disaster news and analysis, including numbers of affected people, building damage, and economic losses. An example is the \textit{Disaster Alert} app (see Figure \ref{recovery}b).
    
    \item \textbf{Share Experiences}: Allowed users to notify others when they experience a disaster, such as the “I Felt It” button in the \textit{Volcanoes \& Earthquakes} app (see Figure \ref{recovery}c).
    
    \item \textbf{Safety Confirmation}: Enabled users to confirm their safety to friends or family through location sharing or short messages within the app, e.g., the \textit{National Evacuation Centre Guide} app (see Figure \ref{recovery}d).
    
    \item \textbf{Find Shelter}: Helped users locate nearby shelters after a disaster. An example is the \textit{Yurekuru Call} app (see Figure \ref{recovery}e).
\end{enumerate}

Overall, the results indicate that disaster mobile apps strongly emphasised response-related features such as real-time alerts and disaster maps, while features supporting preparation and recovery were less commonly implemented.

\subsection{Topics Identified via Topic Modelling (RQ2)}

\begin{figure*}
	\centering
	\includegraphics[width=0.45\textwidth]{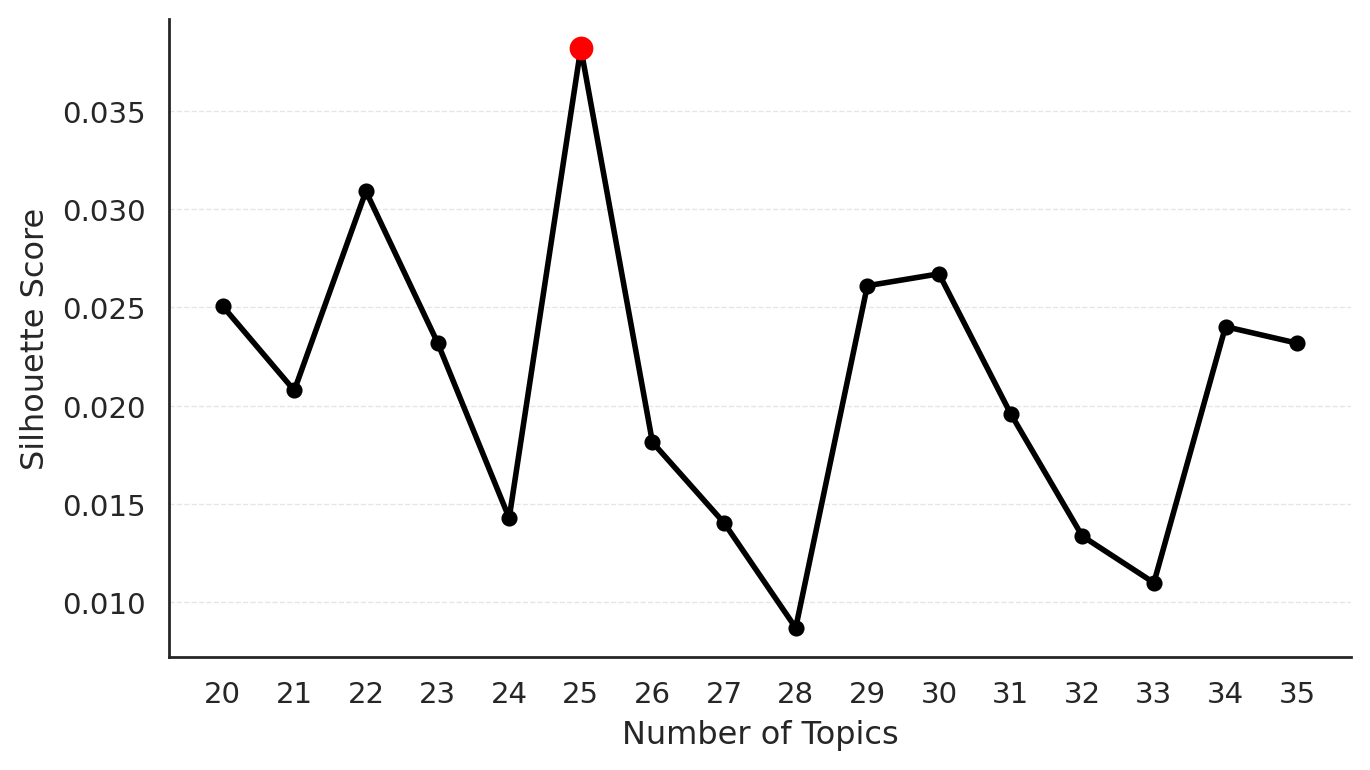}
	\caption{Silhouette score for different numbers of topics used in the clustering analysis.}
	\label{fig:silhouette}
\end{figure*}

A total of 48,621 reviews were analysed using BERTopic to identify the main topics discussed by users. The optimal number of clusters was determined using the Silhouette score, with higher values indicating better clustering. As shown in Figure~\ref{fig:silhouette}, the optimal number of clusters is 25.

Of the 48,621 reviews, 34,664 were assigned to one of the 25 clusters, while the remaining 13,957 reviews were considered outliers. As explained in Section 3.4, to further improve cluster quality, a two-stage clustering approach was applied to detect and remove hidden outliers within the initial clusters. From the 34,664 reviews in the first stage, 26,997 additional outliers were identified and removed, leaving 7,667 clean reviews. After outlier removal, semantically similar clusters were merged. For example, the \textit{Travelling} and \textit{Language} clusters were combined because, in topics related to travellers, the discussion mostly focuses on language barriers when using apps that do not offer English options. Similarly, the \textit{Setting Option} and \textit{App Configuration} clusters were merged, as they refer to the same functionality. Finally, \textit{Navigation} and \textit{User Experience} clusters were combined, since navigation is an integral part of the overall user experience. After these refinements, the final set comprises 22 clusters, as summarised in Table~\ref{tab:topic_distribution}. The table is sorted by discussion frequency, highlighting the topics most frequently mentioned by users. 

\begin{table}[H]
\centering
\caption{Topic Distribution with IDs, Keywords, and Total Reviews}
{\color{black}
\label{tab:topic_distribution}
\scriptsize
\begin{tabular}{clp{5cm}c}
\toprule
\textbf{ID} & \textbf{Topic} & \textbf{Keywords} & \textbf{Total Reviews} \\
\midrule
1  & Alert Functionality       & notifications, timeliness, frequency, relevance, duplicates & 2,284 \\
2  & Advertising and Monetisation & ads, pop-ups, subscriptions, purchases, intrusive & 943 \\
3  & App Crashes                & freezing, closing, launch failure, instability, unresponsive & 611 \\
4  & Keeping Informed            & real-time, tracking, maps, historical, updates & 561 \\
5  & App Reliability             & accuracy, trustworthiness, outdated, timely, dependable & 428 \\
6  & Location Functionality      & GPS, location saving, manual entry, detection, loss & 389 \\
7  & Map Functionality           & interactivity, accuracy, clarity, usability, visualisation & 324 \\
8  & Alert Sound Issues          & notification, audible, volume, override, silent & 324 \\
9  & Battery Consumption         & battery drain, background, wakelocks, foreground, power & 289 \\
10 & User Experience             & ease of use, navigation, intuitive, accessibility, efficiency & 255 \\
11 & User Interface              & design, dark mode, readability, contrast, menus & 248 \\
12 & App Redesign Regression     & updates, redesign, reduced functionality, slower, complexity & 218 \\
13 & App Customisation and Settings & personalisation, filters, preferences, retention, usability & 158 \\
14 & Watch Zone                  & monitoring, radius, naming, customisation, map & 121 \\
15 & Account Registration and Login Issues & registration, login, verification, password, account & 119 \\
16 & Lag Issues                  & slow, unresponsive, delayed, sluggish, navigation & 88 \\
17 & Device Compatibility        & OS, phone models, display, hardware, functionality & 81 \\
18 & Server Connectivity         & connection, downtime, interrupted, network, reliability & 72 \\
19 & Shelter Information         & shelter, accuracy, accessibility, facilities, map & 68 \\
20 & Privacy and Data Access Concerns & permissions, data, personal, camera, contacts & 47 \\
21 & Regional Coverage, National Integration & geographic, cross-region, unified, boundaries, local & 30 \\
22 & English Language Support    & translation, clarity, comprehension, alerts, foreign & 12 \\
\bottomrule
\end{tabular}
}
\end{table}

The results revealed that user discussions focused on several key aspects of disaster mobile apps, including alert functionality, technical reliability, application usability, and information accessibility. Among the most frequently discussed topics were alert notifications, app usability, and the reliability of disaster information provided by the applications. Several topics highlighted technical issues users experienced when interacting with disaster mobile apps. These included application crashes, server connectivity issues, excessive battery consumption, and lag. Such issues show the importance of technical stability in disaster mobile apps, where users rely on the app to deliver timely, accurate information during emergencies. Other topics highlighted user concerns related to application features and configuration. These included map functionality, alert sound settings, and watch-zone configuration. These topics suggest that users frequently engage with features that allow them to customise alerts and monitor disaster events within specific geographic areas. In addition to technical and functional aspects, some topics reflected broader usability and accessibility concerns. For example, several reviews discussed language support and regional coverage, particularly for users travelling abroad or residing in multilingual regions. 

After identifying the topics, we qualitatively analysed the reviews associated with each topic to better understand the specific issues and experiences reported by users. The findings for each topic are presented below.

\subsubsection{Reviews Under Each Topic}

\noindent \textbf{Alert Functionality: }
This topic describes user feedback related to alert behaviour in disaster mobile apps. In some apps, users report not receiving alerts for relevant events, including incidents within configured areas or of expected severity, while in other apps, alerts are described as frequent and repetitive, sometimes referring to the same incident multiple times. Users also note that alerts often arrive late, after an event has already occurred or has been resolved.

\noindent \textbf{Advertising and Monetization:}
This topic captures user feedback about intrusive ads, unexpected in-app purchases, and apps that were once free becoming paid. Users report pop-ups, full-screen ads, and subscription prompts disrupting usage. Some users note that apps are more enjoyable when ads are minimal or removed through paid upgrades.

\noindent \textbf{App Crashes:}
This focuses on user reports of application crashes and freezing. Reviews describe the app closing unexpectedly at launch or shortly after, and freezing during basic interactions, such as selecting menus or viewing content, before closing. These issues are reported to occur repeatedly and prevent users from accessing the application.

\noindent \textbf{Keeping Informed:}
This topic mostly reflects user feedback expressing satisfaction with applications that help them stay informed about ongoing events worldwide. Reviews indicate that users value features that allow them to keep track of what is happening through continuously updated information, including real-time disaster tracking, event maps, and lists of recently occurred incidents. Users also appreciate access to historical records that allow them to review past events and understand patterns over time.

\noindent \textbf{App Reliability:}
This topic captures feedback on the trustworthiness. Users report that information can be outdated, inaccurate, or unclear, making it unreliable. On the other hand, some apps are praised for providing accurate, timely, and detailed information, helping users make informed decisions and stay prepared during emergencies.

\noindent \textbf{Location Functionality:}
This topic reflects user feedback reporting problems with setting, saving, and detecting locations. Users report GPS errors, inability to manually enter locations, crashes when adding locations, and loss of saved locations after updates.

\noindent \textbf{Map Functionality:}
This topic reflects user feedback on map features. Users praise the accuracy, clarity, and interactivity of maps, which help track earthquakes and view additional information. Some report issues with map size, legends, detail, and occasional crashes. Overall, maps are highly valued, but usability can be improved.

\noindent \textbf{Alert Sound Issues:}
This topic captures user feedback related to notification sounds. Reviews note that some apps do not produce audible alerts, leading users to miss notifications. In other apps, alerts are described as excessively loud or difficult to turn off, and in some cases, override device settings such as silent or do not disturb modes.

\noindent \textbf{Battery Consumption:}
This topic reflects extensive user complaints about excessive battery usage caused by the app. Reviews frequently report rapid battery drain even when the app is running in the background, disabled, or not actively used. Many users describe continuous foreground activity, persistent background processes, and wakelocks that prevent devices from entering sleep mode. As a result, users often uninstall the app, expressing concern that an emergency application that significantly reduces battery life undermines its usefulness during critical situations.

\noindent \textbf{User Experience:}
This cluster contains user feedback focused on ease of use, navigation, and accuracy across different apps. Users highlight that some apps are intuitive, straightforward, and user-friendly. In addition, some users note quick access to information, clear interfaces, and effective navigation. However, other users report minor difficulties with some apps, particularly for seniors or less tech-savvy users.

\noindent \textbf{User Interface:}
This feedback focused on the user interface, visual design, and dark mode across different apps. Users frequently request a dark mode or night theme to reduce eye strain and conserve battery life. Some users report issues with dark mode implementation, such as illegible text, poor contrast, or inconsistent behaviour across devices. In addition, users discuss areas needing improvement, such as clarity of menus and support for data visualisation.

\noindent \textbf{App Redesign Regression:}
This cluster captures user feedback comparing the current version of the application with a previous version that users perceived as more reliable and easier to use. Reviews frequently describe dissatisfaction following updates or redesigns, noting reduced functionality, slower response, unclear navigation, missing information, and increased difficulty accessing core features. Many users state a preference for the earlier version, reporting that the newer version introduces complexity without clear benefits and, in some cases, limits access to alerts or critical information that was previously available.

\noindent \textbf{App Customization and Settings:}
This topic captures user feedback about configuring disaster mobile apps to their preferences. Users report difficulties in adjusting, saving, or applying filters and other settings. Problems include settings not being retained, limited filter options, unclear controls, or restrictions behind paid upgrades. While some apps provide useful filtering and customization features, inconsistent behavior and usability issues hinder users from fully personalizing the app experience.

\noindent \textbf{Watch Zone:}
This cluster contains user feedback on watch zone functionality. Users report difficulties with creating, saving, and editing watch zones, including moving the epicenter, adjusting the radius, naming zones, and interacting with the map. The minimum radius is often considered too large for precise monitoring. Despite these challenges, users value watch zones for monitoring emergencies and appreciate the ability to customize alerts for specific areas.

\noindent \textbf{Account Registration and Login Issues:}
This topic contains user complaints about problems creating accounts, logging in, and managing passwords. Users report failed registrations, missing or expired verification codes, inability to reset passwords, and forced account creation even for basic app use. Many describe frustration with complex or broken registration flows, errors across devices, and issues with family or shared account features.

\noindent \textbf{Lag Issues:}
This topic captures user feedback related to poor application performance, particularly lag, slow loading times, and unresponsive interactions. Reviews frequently describe sluggish navigation, delayed screen transitions, and difficulty using core features such as maps, lists, and scrolling, sometimes to the point where the app becomes unusable. Users report that these issues occur across different devices, operating systems, and network conditions, and in some cases worsen after updates.

\noindent \textbf{Device Compatibility:}
This topic captures user feedback related to how the app behaves across different devices and operating system versions. Many reviews report that the app works well on certain phone models and tablets, while others note partial functionality, display issues, glitches, or complete incompatibility on newer or specific devices. Users frequently mention differences in performance after device upgrades or app updates, indicating inconsistent compatibility across hardware and Android versions.

\noindent \textbf{Server Connectivity:}
This topic captures user feedback about connection issues in disaster mobile apps. Users report frequent difficulties connecting to servers, receiving error messages, or experiencing interrupted service. These connectivity problems prevent access to critical updates and reduce the app's usability during emergencies, highlighting the need for more reliable server infrastructure and maintenance.

\noindent \textbf{Shelter Information:}
This topic captures user feedback about locating emergency shelters through disaster mobile apps. Users report difficulties with inaccurate, outdated, or sparse shelter information. Common issues include shelters being shown far from the user, incorrect locations, lack of local coverage, slow maps, or insufficient details about facilities and accessibility. While shelter listings are appreciated, users highlight the need for reliable, up-to-date, and easily navigable information to make the app truly useful in emergencies.

\noindent \textbf{Privacy and Data Access Concerns:}
This topic summarizes user feedback related to privacy and data access permissions requested by the application. Reviews indicate that apps request access to photos, contacts, camera, microphone, and device accounts, which users consider inappropriate for disaster related functionality. Permissions are often accompanied by unclear explanations, contributing to uncertainty and concern about data access and use.

\noindent \textbf{Regional Coverage and National Integration:}
This topic captures user feedback related to the geographic scope of disaster information provided by the app. While users value accurate and detailed coverage within supported regions, many report frustration when events in nearby or neighbouring areas are missing, outdated, or inaccessible. Reviews frequently highlight difficulties for users who travel across regions or live near boundaries, where switching between multiple apps or sources is required to obtain complete information. These limitations reduce the app’s usefulness in large scale or cross regional events and lead users to request broader, unified coverage within a single application.

\noindent \textbf{English Language Support:}
This topic reflects user feedback requesting proper English translation in the app. Users report that alerts, maps, and other information are sometimes displayed in foreign languages, making them difficult to understand, especially for travellers or foreign residents.

\subsection{Sentiment Analysis of Disaster Mobile Apps Reviews (RQ3)}
To better understand the challenges users experience when using disaster mobile apps, we analysed the sentiment expressed in user reviews using the VADER sentiment analysis method described in Subsection~\ref{subsec:sentiment_analysis}. The compound sentiment score produced by VADER was used to classify reviews as positive, neutral, or negative. For user ratings, reviews were categorised as follows: ratings of 1--2 are considered negative, 3 as neutral, and 4--5 as positive.

\subsubsection{Distribution of Sentiment}

{Table~\ref{tab_sentiment_distribution} summarises the sentiment distribution of the app reviews based on both user ratings and text-derived sentiment scores. According to user ratings, 49.11\% of reviews are positive, 40.05\% negative, and 10.84\% neutral. Text-based sentiment analysis yields a slightly different distribution: 42.77\% positive, 44.67\% negative, and 12.56\% neutral. This discrepancy reflects the tendency for users to provide high ratings despite highlighting multiple issues, as discussed in the previous section. Overall, the analysis reveals a substantial proportion of negative feedback. In the following subsection, we examine the topics associated with negative reviews to identify areas where developers can prioritise improvements.

\begin{table}[h]
\caption{\small Sentiment Analysis by User Rating and Sentiment Score}
{\color{black}
\label{tab_sentiment_distribution}
\resizebox{0.6\columnwidth}{!}{
\scriptsize
\begin{tabular}{@{}ll@{}}
\toprule
\textbf{Sentiment} & \textbf{Number of Reviews [\%]}  \\ 
\midrule
User Rating \\
\midrule
Negative & \mybarhhigh{40.05} {3071 [40.05\%]}  \\
Neutral & \mybarhhigh {10.84} {831 [10.84\%]} \\
Positive & \mybarhhigh {49.11} {3765 [49.11\%]}  \\
\midrule
Sentiment Score \\
\midrule
Negative & \mybarhhigh{44.67} {3425 [44.67\%]}  \\
Neutral & \mybarhhigh {12.56} {963 [12.56\%]} \\
Positive & \mybarhhigh {42.77} {3279 [42.77\%]}  \\
\bottomrule
\end{tabular}
}
}
\end{table}

\subsubsection{Topics with Lowest Sentiment and Ratings}

Table~\ref{tab:topic_sentiment} presents both user ratings and VADER compound scores for each topic. Higher values indicate more positive user responses. The table also includes the percentage of negative reviews based on user ratings and sentiment analysis, where higher percentages reflect more negative perceptions. To facilitate interpretation, the top three lowest user ratings and compound scores are highlighted in red, along with the top three highest percentages of negative reviews and negative sentiment. Conversely, the top three highest user ratings and compound scores, as well as the lowest percentages of negative reviews and sentiment, are highlighted in green.

From the table, it is evident that several topics receive a high proportion of negative feedback from both user ratings and sentiment analysis, indicating areas that require attention. Conversely, some topics consistently receive positive feedback and should be maintained. Based on this, we categorise the topics into two groups: those that should be maintained for quality and those with urgent issues that need to be addressed.

\noindent \textbf{Topics related to quality:}  
\begin{itemize}
    \item \textit{Keeping Informed:} Users consistently value real-time updates, event maps, and historical records that allow them to monitor ongoing disasters and understand trends. Maintaining these features, or adding them if the app currently lacks them, will enhance user satisfaction.
    \item \textit{Map Functionality:} Interactive and accurate maps facilitate understanding of events. Continuing to optimise map usability, including responsiveness and clarity, will reinforce this key strength
    \item \textit{User Experience:} Intuitive navigation, clear interfaces, and easy access to information contribute to positive user perception. Retaining these design elements supports app accessibility across diverse user groups
\end{itemize}

\noindent \textbf{Topics with issues requiring urgent resolution:}  
\begin{itemize}
    \item \textit{App Crashes:} Frequent crashes and freezing prevent access to critical features. Immediate investigation of app stability and memory management is needed to restore usability. 
    \item \textit{Watch Zone:} Users encounter difficulties creating, editing, and saving watch zones, including radius adjustments and map interaction. Redesigning the watch zone interface to allow precise placement, variable radii, and easier naming will improve functionality.  
    \item \textit{Account Registration and Login Issues:} Registration failures, password errors, and forced account creation hinder user access. Simplifying account setup, fixing verification flows, and allowing optional profiles for core features will reduce friction.  
    \item \textit{Lag Issues:} Slow loading, delayed navigation, and unresponsive features degrade the experience. Optimising performance across devices and network conditions, including updates, is critical.  
    \item \textit{Server Connectivity:} Frequent server errors and disconnections limit access to essential updates. Strengthening server reliability and implementing redundancy is necessary for uninterrupted service.  
    \item \textit{Shelter Information:} Inaccurate, outdated, or incomplete shelter data reduces usefulness. Updating shelter databases, improving geolocation accuracy, and providing detailed facility information will enhance emergency preparedness.
    \item \textit{Privacy and Data Access Concerns:} Requests for unnecessary permissions create user distrust. Minimising data access, clarifying permission purposes, and following privacy-by-design principles will improve confidence and adoption.
\end{itemize}

\begin{table}[ht]
\centering
\small
\caption{Summary of Topic Ratings and Sentiment.}
{\color{black}
\scriptsize
\begin{tabular}{c p{3.5cm} c c c c}
\toprule
\textbf{ID} & \textbf{Topic} & \textbf{Avg User Rating} & \textbf{Avg Compound Score} & \textbf{Neg. Review \%} & \textbf{Neg. Sentiment \%} \\
\midrule
1 & Alert Functionality & 3.22 & 0.137 & 39.06 & 50.50 \\
2 & Advertising and Monetisation & 2.75 & 0.128 & 52.17 & 50.69 \\
3 & App Crashes & \cellcolor{red!40}2.00 & 0.100 & \cellcolor{red!40}72.01 & 59.74 \\
4 & Keeping Informed & \cellcolor{green!40}4.64 & \cellcolor{green!40}0.561 & \cellcolor{green!40}6.42 & \cellcolor{green!40}13.19 \\
5 & App Reliability & \cellcolor{green!40}4.54 & \cellcolor{green!40}0.390 & 7.01 & 30.61 \\
6 & Location Functionality & 2.13 & 0.049 & 64.52 & 61.44 \\
7 & Map Functionality & 4.00 & \cellcolor{green!40}0.500 & 16.67 & \cellcolor{green!40}21.60 \\
8 & Alert Sound Issues & 3.29 & 0.146 & 28.09 & 47.53 \\
9 & Battery Consumption & 2.38 & 0.088 & 56.40 & 57.44 \\
10 & User Experience & \cellcolor{green!40}4.62 & \cellcolor{green!40}0.544 & \cellcolor{green!40}5.49 & \cellcolor{green!40}8.63 \\
11 & User Interface & 3.89 & 0.434 & 17.74 & 23.39 \\
12 & App Redesign Regression & 2.39 & 0.245 & 61.47 & 34.86 \\
13 & App Customisation and Settings & 3.28 & 0.225 & 34.18 & 45.57 \\
14 & Watch Zone & 2.08 & 0.116 & \cellcolor{red!40}69.42 & 51.24 \\
15 & Account Registration and Login Issues & \cellcolor{red!40}1.62 & \cellcolor{red!40}-0.035 & \cellcolor{red!40}84.03 & 64.71 \\
16 & Lag Issues & 2.44 & \cellcolor{red!40}-0.079 & 55.68 & \cellcolor{red!40}64.77 \\
17 & Device Compatibility & 3.93 & 0.329 & 20.99 & 33.33 \\
18 & Server Connectivity & 2.08 & 0.015 & 63.89 & \cellcolor{red!40}70.83 \\
19 & Shelter Information & 2.82 & \cellcolor{red!40}0.004 & 48.53 & 63.24 \\
20 & Privacy and Data Access Concerns & \cellcolor{red!40}2.04 & 0.032 & 63.83 & \cellcolor{red!40}74.47 \\
21 & Regional Coverage and National Integration & 3.23 & 0.259 & 40.00 & 36.67 \\
22 & English Language Support & 3.00 & 0.084 & 50.00 & 41.67 \\

\bottomrule
\end{tabular}
\label{tab:topic_sentiment}
}
\end{table}

In summary, the findings indicate that while disaster mobile apps commonly provide core response features, persistent challenges remain in some key areas with issues. These areas represent key opportunities for improving the design and implementation of future disaster mobile apps.

\section{Discussion}

The results of this study provide insights into the current capabilities and limitations of disaster mobile apps. By analysing app features, user reviews, and sentiment patterns, several key themes emerged regarding how these applications support disaster risk reduction and where improvements are needed.

\subsection{Feature Imbalance Across the Disaster Lifecycle}
Our feature analysis revealed that most disaster mobile apps primarily focus on response-related functionalities, such as real-time alerts and disaster maps. While these features play a critical role during emergency situations, the results suggest that other stages of the disaster lifecycle—particularly preparation and recovery—are less well supported. Preparation-related features, such as disaster forecasting tools and preparedness guidance, were present in relatively few applications. Similarly, recovery-oriented functionalities, including shelter information and safety confirmation, were limited across the analysed apps. These findings indicate that many disaster mobile apps prioritise immediate situational awareness during disaster events, while providing less support for long-term preparedness and post-disaster recovery. This imbalance may be influenced by several factors. First, disaster management practices have traditionally prioritised reactive measures, such as response, over proactive measures, such as preparedness and mitigation \cite{navarro2022assessment}. Second, chronic underinvestment in disaster preparedness may contribute to a shortage of digital tools focused on long-term risk reduction, while immediate, high-visibility response technologies receive greater support \cite{chan2023alert}. Third, the distributed nature of the disaster management landscape, where different agencies are responsible for response and recovery activities, may lead to fragmented support across lifecycle stages \cite{abbas2025exploring}. In addition, many disaster mobile apps are developed by agencies primarily responsible for the response phase, which may further contribute to the emphasis on response-oriented functionalities.

Expanding app capabilities to cover the full disaster lifecycle could improve the effectiveness of mobile apps as disaster risk reduction tools. For example, early warning alerts and preparedness guidance can support preventive action before disasters occur, while recovery-related features can assist affected communities in locating assistance and reconnecting with family members after an event. In addition, greater coordination among agencies responsible for different lifecycle stages may help ensure that consistent, comprehensive information is delivered through mobile apps. Such integration can enhance the role of disaster mobile apps in supporting communities before, during, and after disaster events.

\subsection{Internationalisation and Language Barriers}
Language accessibility emerged as an important usability challenge for many disaster mobile apps. Our analysis showed that while global and regional applications often provide English-language support, many national apps developed in non-English-speaking countries offer limited multilingual options. This limitation may create barriers for tourists, temporary residents, and international communities who rely on disaster information while travelling or living abroad. Often, this is accounted for by the "Utilitarian" approach, where agencies default to sending the same message in the area's dominant language to ensure all information reaches the majority as quickly as possible, which can leave non-native speakers at risk \cite{uekusa2023preparing}. Improving internationalisation by providing multilingual interfaces and ensuring translation quality could significantly enhance the accessibility and usability of disaster mobile apps. Such improvements would allow disaster information to reach broader audiences, particularly in regions with high levels of tourism or linguistic diversity.

\subsection{Balancing Monetisation and Public Safety}
Another important consideration concerns the monetisation strategies adopted by some disaster mobile apps. Although many apps were freely available, some included advertisements or paid features. User reviews suggested that intrusive advertisements may negatively affect the user experience, particularly during time-critical situations when users need to access information quickly. Advertisements that appear during emergency notifications or critical updates may distract users or delay access to important disaster information. In addition, applications that require payment may experience lower adoption rates, limiting their reach during disasters \cite{tanaka2025pays}. Such models may also introduce barriers for individuals in lower socioeconomic settings, potentially restricting equitable access to critical hazard information \cite{tan2020usability}. Developers should therefore carefully consider monetisation strategies for disaster mobile apps. One possible approach is to disable advertisements during active disaster events while maintaining alternative revenue models such as optional subscription-based services. More broadly, aligning monetisation strategies with the public safety objectives of disaster communication systems is critical for ensuring that mobile apps effectively serve diverse communities during emergencies.

\subsection{Technical Reliability Challenges}
The sentiment analysis revealed that several topics associated with technical reliability generated high levels of negative sentiment. These included application crashes, connectivity problems, battery consumption issues, and sign-up or login failures. Such issues represent significant usability barriers because they may prevent users from accessing disaster information when it is most needed. In the context of emergency situations, technical failures may undermine user trust in the application and discourage continued use. When users experience repeated failures—such as alerts arriving after an incident has already occurred—they may reduce their reliance on official apps and instead turn to alternative platforms such as social media. Therefore, improving application stability through rigorous testing, optimised background processes, and robust network handling mechanisms is essential. Ensuring reliable performance under high-load conditions, such as during major disaster events when many users access the app simultaneously, is particularly important.

\subsection{Usability and Inclusiveness}
User reviews revealed several accessibility and inclusivity challenges. For example, some users reported difficulties interpreting colour-coded maps due to colour vision impairments, while others noted that small font sizes made information difficult to read. Similarly, some users indicated that alert sounds were either excessively loud or too quiet, affecting individuals with hearing sensitivities. These challenges reflect a broader tendency within the disaster domain to treat inclusivity as a “last-mile” concern, addressed after systems are developed, rather than as a “first-mile” design consideration \cite{faal2022overcoming}. However, inclusivity can be argued to be most effective when integrated from the outset. Designing systems to accommodate specific user groups often leads to broader usability benefits for all users. For instance, subtitles and closed captions—originally designed for deaf and hard-of-hearing individuals—are now widely used in noisy environments or when audio cannot be enabled.

These findings suggest that disaster mobile apps should adopt more inclusive design practices. This aligns with broader disaster risk reduction (DRR) literature, which emphasises the importance of accessible hazard communication \cite{stjernholm2025active}. Given the diverse user base of disaster mobile apps, accessibility should be treated as a core design requirement rather than an optional enhancement. Practical improvements could include adjustable font sizes, high-contrast map visualisations, customisable alert settings, and better compatibility with assistive technologies. Such features can improve usability for diverse user groups and ensure that critical information remains accessible during emergency situations. More broadly, incorporating accessibility considerations early in the design process can help ensure that disaster information systems are usable by a wider range of individuals, including people with disabilities, older adults, and users in low-connectivity environments. Addressing these accessibility gaps can improve comprehension, reduce cognitive load, and strengthen user trust in hazard information systems, all of which are essential components of a safety-critical system such as a disaster early warning system.

\section{Recommendations}

Disaster mobile apps operate within a socio-technical ecosystem involving both software developers and emergency management agencies. While developers are responsible for the technical design and implementation of mobile apps, the warning messages, hazard information, and other disaster-related content are typically provided by emergency agencies. Therefore, addressing the challenges identified in this study requires coordinated improvements from both groups. Technical enhancements alone may not fully resolve users' reported issues if the underlying disaster communication practices remain unchanged. For this reason, we provide recommendations directed at both developers and emergency agencies. In the following subsections, \faLaptop\ indicates recommendations primarily for developers, while \faAmbulance\ indicates recommendations for emergency management agencies. 

\subsection{Develop Lifecycle-Oriented Disaster Mobile Apps}
Our analysis showed that most disaster mobile apps prioritise response-related features such as real-time alerts and disaster maps, while providing limited support for preparedness and recovery. To address this limitation, 

\textbf{\faLaptop:}
Developers should design applications that support the full disaster lifecycle by incorporating features that facilitate preparedness, response, and recovery. This may include functionalities such as preparedness guidance, evacuation planning tools, hazard awareness information, and recovery-related services such as shelter directories and safety confirmation mechanisms. Designing applications with modular, flexible structures can also enable additional lifecycle features to be integrated as new disaster communication needs emerge.

\textbf{\faAmbulance:}
Emergency agencies can support lifecycle-oriented applications by providing reliable information and resources relevant to different disaster phases. This includes preparedness materials, early warning information, evacuation centre data, and post-disaster recovery guidance that can be integrated into mobile apps. Ensuring that such information remains regularly updated and accessible can significantly enhance the usefulness of disaster mobile apps across all stages of disaster management.

\subsection{Strengthen Internationalisation and Multilingual Support}
Language barriers can significantly limit the accessibility of disaster mobile apps, particularly for tourists, temporary residents, and multilingual communities. In disaster situations, the inability to understand warning messages may delay protective actions and increase risk. To address this, 

\textbf{\faLaptop:}
Developers should implement multilingual interfaces that allow users to easily switch languages within the application. Supporting widely used languages, such as English, alongside local languages can improve accessibility for international users. Opportunities exist here to use the latest AI technologies, including Large Language Models (LLMs), to assist with the translation. However, in using such methods, developers should always ensure that translation quality is carefully validated, as inaccurate translations may lead to misunderstandings of hazard information. Additional design features, such as enabling users to copy text for use with external translation tools and providing simple, easy-to-understand message structures, can further enhance usability.

\textbf{\faAmbulance:}
Emergency agencies can further support multilingual accessibility by ensuring that official warning messages and hazard information are available in multiple languages and can be disseminated through mobile apps. Rather than relying solely on automatic translations provided by mobile platforms, agencies can develop verified multilingual warning templates and terminology guidelines that ensure the accuracy and consistency of disaster messaging. This may involve preparing pre-translated warning templates, collaborating with professional translation services, or establishing multilingual communication protocols for emergency alerts. Ensuring that translated warnings are delivered alongside primary alerts can help diverse populations receive timely and understandable hazard information while preserving the accuracy and authenticity of official communications.

\subsection{Adopt User-Centred and Inclusive Design Practices}
Beyond language accessibility, disaster mobile apps must also consider the diverse capabilities and needs of their users. Disaster situations often require individuals to quickly interpret warning information and make immediate decisions, often under stressful conditions. Our analysis of user reviews revealed several usability and accessibility challenges related to interface design, visualisation, and alert mechanisms. To address this,

\textbf{\faLaptop:}
Developers should adopt user-centred and inclusive design principles when developing disaster mobile apps. When inclusivity is treated as a ``First-mile" consideration by developers, this helps to improve accessibility of systems for all users. Interfaces should prioritise simplicity and readability to ensure that users can quickly interpret hazard information. Practical measures include: 

\begin{itemize}
\item Providing adjustable text sizes and readable typography
\item Using high-contrast visualisations and colour-blind friendly palettes for maps and alerts
\item Simplifying hazard maps and visual hierarchies to reduce cognitive load
\item Allowing users to customise alert settings, including sound levels, vibration patterns, and notification types
\item Supporting offline functionality or low-data modes to ensure access in low-connectivity environments
\item Ensuring compatibility across a wide range of devices and operating systems to broaden access
\end{itemize}

\textbf{\faAmbulance:}
Emergency agencies play an important role in ensuring that the information delivered through these applications is accessible and understandable. Agencies should provide warning messages that use clear language, consistent terminology, and simple message structures. Designing warnings that avoid unnecessary technical terminology can help ensure hazard information is quickly interpretable by diverse user groups. Agencies may also benefit from engaging with diverse communities when designing communication strategies, ensuring that disaster messaging reflects the needs of different populations.

\subsection{Improve Technical Reliability and Performance Stability}
Technical reliability emerged as one of the most prominent concerns expressed in user reviews. To address this developers and emergency agencies can take the following actions. 

\textbf{\faLaptop:}
Developers should prioritise the reliability and performance stability of disaster mobile apps. This includes ensuring that applications are thoroughly tested under high-demand scenarios that simulate disaster conditions, where large numbers of users may simultaneously access alerts and information. Developers should also optimise background processes and location services to minimise battery consumption, as disaster mobile apps are often required to operate continuously during emergencies. In addition, implementing robust error-handling mechanisms can help prevent application crashes and ensure stable performance. Developers should also design systems that deliver critical alerts even when network connectivity is limited, for example, by supporting cached alerts or low-bandwidth communication mechanisms. Simplifying authentication processes or allowing guest access to essential information can further help ensure that users can access critical hazard information quickly during emergency situations.

\textbf{\faAmbulance:}
Emergency agencies should ensure that the infrastructure supporting disaster mobile apps can handle high volumes of user traffic during emergency events. Maintaining reliable data services, ensuring the timely dissemination of warning messages, and coordinating closely with application developers can help ensure that alerts and hazard information are delivered consistently during disaster situations. Strengthening the reliability of these supporting systems can significantly improve the effectiveness of disaster mobile apps as communication tools during emergencies.

\subsection{Adopt Responsible Monetisation Strategies}
Some disaster mobile apps incorporate advertisements or subscription models that may affect user experience during emergencies. This can be addressed by,

\textbf{\faLaptop:}
Developers should ensure that monetisation mechanisms do not interfere with the delivery of critical disaster information. In particular, advertisements should not obscure alerts, hazard maps, or evacuation instructions. Developers can implement safeguards such as automatically disabling advertisements when emergency alerts are active or ensuring that critical warnings are always prioritised in the user interface.

\textbf{\faAmbulance:}
Emergency agencies should encourage monetisation practices that prioritise public safety and accessibility when collaborating with developers or third-party platforms. Agencies may also establish guidelines to ensure that disaster communication tools remain focused on delivering timely and reliable hazard information without unnecessary distractions during emergencies. For instance, when disaster mobile apps include advertisements or paid features, essential warning information should remain freely accessible to all users.

\subsection{Integrate Continuous User Feedback}
User reviews provide valuable insights into real-world experiences with disaster mobile apps. In this study, topic modelling and sentiment analysis revealed recurring concerns related to alert functionality, technical reliability, usability, and information accessibility. These findings demonstrate that user feedback can serve as an important source of evidence for identifying system limitations and areas for improvement. Therefore, the following measures can be adopted to ensure that user feedback is systematically considered.

\textbf{\faLaptop:}
Developers should adopt systematic approaches to monitoring and analysing user feedback from app stores and other platforms. Rather than relying on ad hoc review, developers can use analytical techniques such as topic modelling and sentiment analysis to identify recurring issues, prioritise frequently reported problems, and track changes in user satisfaction over time. Integrating these insights into iterative development cycles can help ensure that applications are continuously refined based on real user needs.

\textbf{\faAmbulance:}
Emergency agencies can also explore accessing and using the review feedback to evaluate the effectiveness of disaster communication delivered through mobile apps. Analysing user reviews can help agencies identify issues related to message clarity, alert timing, and information relevance. These insights can inform improvements in warning content, communication strategies, and the overall design of disaster messaging. Incorporating user feedback into ongoing evaluation processes can help ensure that disaster communication systems remain responsive to diverse communities' needs.

\subsection{Promote Cross-Sector Collaboration}
The findings of this study highlight that disaster mobile apps are not purely technical systems, but socio-technical systems that rely on coordination between developers and emergency management agencies. Many of the challenges identified, including issues related to usability, reliability, and information clarity, arise from the interaction between technical design and disaster communication practices. Therefore, to ensure better collaboration among these parties, the following actions can be taken.

\textbf{\faLaptop:}
Developers should actively engage with emergency management agencies to better understand operational requirements, disaster response workflows, and communication needs. Incorporating domain knowledge into the design process can help ensure that applications align with real-world emergency scenarios and user expectations. In addition, developers should adopt more effective requirements elicitation and validation practices to capture the complex contextual realities of disaster environments. This may include the design and adoption of domain-specific modelling approaches, such as Domain-Specific Visual Languages (DSVLs), to support clearer communication of requirements and facilitate validation with emergency agencies. Such approaches can help ensure that system functionalities accurately reflect stakeholder needs and operational contexts.

\textbf{\faAmbulance:}
Emergency agencies should actively participate in the design and development of disaster mobile apps by clearly stating their communication requirements, operational workflows, and information priorities. Providing structured and timely input during requirement elicitation can help ensure that critical contextual information is accurately captured. Agencies should also be involved in validating system functionalities, for example, by reviewing visual representations or prototypes, to confirm that the application aligns with real-world disaster response practices. In addition, engaging with diverse user groups during co-design activities can help ensure that disaster communication strategies reflect the needs of different communities and contexts.

\section{Limitations}

\textbf{Disaster Mobile Apps Dataset:} In this study, we created the disaster mobile apps collection using multiple methods to capture both well-known and lesser-known apps across different countries. These included a literature-based app search, a keyword-based search across popular app distribution platforms, and a country-based search on Google. While this approach enabled a systematic, scalable data collection process, it may have excluded applications that are not publicly listed or region-specific and difficult to access. As a result, the set of analysed applications may not fully represent all disaster communication apps used globally. The second limitation arises from the filtering process we used to narrow down the disaster mobile apps. We applied a false-positive filter to identify apps that did not provide functionalities directly related to natural disaster management. This process risked excluding relevant apps with disaster-related features due to misclassification. To address this issue, we downloaded and tested each app directly to verify its relevance and functionality. Furthermore, all three authors engaged in regular discussions to refine the inclusion criteria, ensuring that the final dataset accurately represents disaster management applications.

\textbf{Review Dataset: }To create this dataset, data scraping was performed using both the Google Play Store and the App Store scrapers. In using these, we employed batch scraping strategies to manage access limitations imposed by Google and Apple. This meant that we paused for 5 seconds after every 500 reviews. But despite these precautions, some reviews may still have been missed if access was denied due to scraping limits.

\textbf{Analysis:} In this study, our analysis relied on a single clustering technique, BERTopic, and a single sentiment analysis method, a rule-based model called VADER. Relying solely on these methods might limit the generalizability of our findings, as different techniques could potentially yield varying results. Additionally, the configuration and fine-tuning of BERTopic might have also influenced the results. The hyperparameters, such as the sentence embedding method, distance metric formula, and minimum topic size, can significantly affect what the model identifies as a topic. Similarly, the configuration of the VADER model, particularly the sentiment thresholds used to classify texts as positive ( 1 >= score >= 0.33), neutral (0.33 > score > -0.33), or negative (-0.33 >= score >= -1), may affect sentiment analysis. Since VADER is rule-based, there are no hyperparameters to tune, but the chosen thresholds are crucial for sentiment classification. These factors suggest caution when interpreting the results. Future studies could benefit from incorporating multiple clustering and sentiment analysis techniques to validate the findings. For sentiment analysis, comparisons between rule-based methods like VADER and machine learning techniques could be particularly illuminating. However, applying machine learning techniques would require access to well-labelled data, which is essential for accurately training and testing the models. 

\textbf{Context:} Finally, the findings of this study may be influenced by the temporal context in which the data were collected. Disaster mobile apps are continuously updated, and user perceptions may change over time as new features are introduced or existing issues are resolved. Therefore, the results should be interpreted as a snapshot of the disaster mobile apps ecosystem during the study period.

Despite these limitations, this study provides a comprehensive analysis of disaster mobile apps and user-reported challenges, offering insights that can inform improvements to future disaster communication systems. To enhance the reliability of our study and ensure reproducibility for future researchers, we have compiled a comprehensive replication package, openly available on author's GitHub repository, which includes the dataset used in our analysis and the source code for the clustering and sentiment analysis models utilised. This repository is designed to facilitate replication of our study by providing step-by-step guidelines for use, thereby minimising the risk of obtaining different results.


\section{Conclusion}
This paper presented a comprehensive analysis of disaster mobile apps by examining both their features and user-reported experiences. By analysing a curated set of disaster mobile apps and a large corpus of user reviews, this study identified key characteristics of existing applications, explored the main topics discussed by users, and examined the challenges users encounter when interacting with these systems. The findings revealed several important insights. Firstly, disaster mobile apps predominantly focus on response-related functionalities, with comparatively limited support for preparedness and recovery stages. Secondly, user reviews highlighted a range of challenges related to technical reliability, usability, accessibility, and multilingual support. These issues can significantly affect how users interpret and act on disaster information, particularly in time-critical, high-stress situations. Thirdly, the analysis demonstrated the value of user reviews as a rich source of feedback for understanding real-world experiences and identifying system limitations. This study contributes to the disaster risk reduction and software engineering literature by providing an integrated perspective on disaster mobile apps, combining feature analysis with large-scale user feedback analysis. The findings emphasise that improving disaster mobile apps requires not only technical enhancements but also improvements in how disaster information is communicated and delivered to diverse user groups. Future work can extend this research in several directions. This includes conducting user studies with diverse communities to better understand accessibility needs, exploring the integration of emerging technologies to support inclusive disaster communication, and investigating how disaster mobile apps can be more effectively integrated within broader early warning systems. Overall, this study highlights the importance of designing reliable and user-centred disaster mobile apps that can effectively support communities before, during, and after disaster events.


\clearpage
\bibliographystyle{elsarticle-num-names}
\bibliography{references_jss}

\clearpage
\appendix

\section*{Appendix A}
\setlength{\tabcolsep}{4pt}
\renewcommand{\arraystretch}{1.1}

{\small
{\color{black} 
\begin{longtable}{
c
P{2.2cm}
P{2.2cm}
c
c
P{2.2cm}
P{2.2cm}
P{2.2cm}
}
\caption{List of Disaster Mobile Applications Included in the Study}
\label{tab:apps}\\
\toprule
ID & App & Source & Downloads & Rating & Pricing & \makecell{Geographical\\Coverage} & Disaster Type \\
\midrule
\endfirsthead

\toprule
ID & App & Source & Downloads & Rating & Pricing & \makecell{Geographical\\Coverage} & Disaster Type \\
\midrule
\endhead

\midrule
\multicolumn{8}{r}{\textit{Continued on next page}}
\endfoot

\bottomrule
\endlastfoot

1 &
\multirow{2}{*}{\parbox[t]{2.2cm}{\raggedright Alabama SAF-T-Net}} &
Apple App Store & 100+ & 3.3 & Free & Local & Multi-Disaster \\
 &  & Google Play Store & 50K+ & 3.3 & Free with Ads & Local & Multi-Disaster \\

2 &
\multirow{2}{*}{\parbox[t]{2.2cm}{\raggedright Alberta Emergency Alert}} &
Apple App Store & 14K+ & 4.7 & Free & Local & Multi-Disaster \\
 &  & Google Play Store & 100K+ & 4.7 & Free & Local & Multi-Disaster \\

3 &
\multirow{2}{*}{\parbox[t]{2.2cm}{\raggedright Alert SA}} &
Apple App Store & 7K+ & 4.7 & Free & Local & Multi-Disaster \\
 &  & Google Play Store & 100K+ & 4.5 & Free & Local & Multi-Disaster \\

4 &
\parbox[t]{2.2cm}{\raggedright Alert2Me Emergency Alerts} &
Google Play Store & 10K+ & 3.4 & Free & Global & Multi-Disaster \\

5 &
\multirow{2}{*}{\parbox[t]{2.2cm}{\raggedright Alertswiss}} &
Apple App Store & 600+ & 3.5 & Free & National & Multi-Disaster \\
 &  & Google Play Store & 500K+ & 3.4 & Free & National & Multi-Disaster \\

6 &
\parbox[t]{2.2cm}{\raggedright Anhaar} &
Google Play Store & 10K+ & 4.7 & Free & National & Multi-Disaster \\

7 &
\parbox[t]{2.2cm}{\raggedright BD 999} &
Google Play Store & 10K+ & 4.5 & Free & National & Multi-Disaster \\

8 &
\multirow{2}{*}{\parbox[t]{2.2cm}{\raggedright Disaster Alert}} &
Apple App Store & 700+ & 4.2 & Free & Global & Multi-Disaster \\
 &  & Google Play Store & 1M+ & 3.6 & Free & Global & Multi-Disaster \\

9 &
\multirow{2}{*}{\parbox[t]{2.2cm}{\raggedright Earthquake}} &
Apple App Store & 3K+ & 4.8 & Free with In-app Purchase & Global & Earthquake \\
 &  & Google Play Store & 100K+ & 4.1 & Free with Ads & Global & Earthquake \\

10 &
\parbox[t]{2.2cm}{\raggedright Earthquake alerts and map} &
Apple App Store & 45K+ & 4.7 & Free with In-app Purchase & Global & Earthquake \\

11 &
\multirow{2}{*}{\parbox[t]{2.2cm}{\raggedright Earthquake + Alerts Map \& Info}} &
Apple App Store & 22K+ & 4.7 & Free with In-app Purchase & Global & Earthquake \\
 &  & Google Play Store & 100K+ & 4.2 & Free with Ads and In-app Purchase & Global & Earthquake \\

12 &
\parbox[t]{2.2cm}{\raggedright Earthquake Alert!} &
Google Play Store & 1M+ & 4.6 & Free & Global & Earthquake \\

13 &
\parbox[t]{2.2cm}{\raggedright Earthquake Alerts \& Tracker} &
Google Play Store & 50K+ & 4.4 & Free with Ads & Global & Earthquake \\

14 &
\parbox[t]{2.2cm}{\raggedright Earthquake App Tracker Map} &
Google Play Store & 500K+ & 4.3 & Free with Ads and In-app Purchase & Global & Earthquake \\

15 &
\multirow{2}{*}{\parbox[t]{2.2cm}{\raggedright Earthquake Network}} &
Apple App Store & 2K+ & 4.6 & Paid with In-app Purchase & Global & Earthquake \\
 &  & Google Play Store & 10M+ & 4.3 & Free with Ads and In-app Purchase & Global & Earthquake \\

16 &
\multirow{2}{*}{\parbox[t]{2.2cm}{\raggedright Earthquake Turkey}} &
Apple App Store & 11K+ & 4.6 & Free with In-app Purchase & National & Earthquake \\
 &  & Google Play Store & 100K+ & 4.6 & Free with In-app Purchase & National & Earthquake \\

17 &
\parbox[t]{2.2cm}{\raggedright Earthquake Quake Tracker} &
Apple App Store & 2K+ & 4.7 & Free with In-app Purchase & Global & Earthquake \\

18 &
\parbox[t]{2.2cm}{\raggedright Earthquake+} &
Apple App Store & 2K+ & 4.6 & Free with In-app Purchase & Global & Earthquake \\

19 &
\parbox[t]{2.2cm}{\raggedright Earthquakes} &
Google Play Store & 500K+ & 3.0 & Free with Ads and In-app Purchase & Global & Earthquake \\

20 &
\parbox[t]{2.2cm}{\raggedright Earthquakes Near Me} &
Google Play Store & 100K+ & 4.5 & Free with Ads & Global & Earthquake \\

21 &
\parbox[t]{2.2cm}{\raggedright Earthquakes Today} &
Google Play Store & 100K+ & 4.3 & Free with Ads and In-app Purchase & Global & Earthquake \\

22 &
\parbox[t]{2.2cm}{\raggedright Earthquakes Tracker} &
Google Play Store & 500K+ & 4.5 & Free with Ads and In-app Purchase & Global & Earthquake \\

23 &
\parbox[t]{2.2cm}{\raggedright Earthquakes Tracker Pro} &
Google Play Store & 5K+ & 4.5 & Paid & Global & Earthquake \\

24 &
\parbox[t]{2.2cm}{\raggedright Emergency} &
Google Play Store & 10K+ & 4.2 & Free & National & Multi-Disaster \\

25 &
\multirow{2}{*}{\parbox[t]{2.2cm}{\raggedright Emergency Ready App}} &
Apple App Store & 20+ & 3.0 & Free & National & Multi-Disaster \\
 &  & Google Play Store & 10K+ & 3.3 & Free & National & Multi-Disaster \\

26 &
\multirow{2}{*}{\parbox[t]{2.2cm}{\raggedright Emergency Severe Weather App}} &
Apple App Store & 4K+ & 4.9 & Free & National & Multi-Disaster \\
 &  & Google Play Store & 100K+ & 4.5 & Free & National & Multi-Disaster \\

27 &
\parbox[t]{2.2cm}{\raggedright EQInfo} &
Google Play Store & 100K+ & 3.0 & Free & Global & Earthquake \\

28 &
\multirow{2}{*}{\parbox[t]{2.2cm}{\raggedright FEMA}} &
Apple App Store & 900+ & 3.7 & Free & National & Multi-Disaster \\
 &  & Google Play Store & 1M+ & 4.4 & Free & National & Multi-Disaster \\

29 &
\multirow{2}{*}{\parbox[t]{2.2cm}{\raggedright Fires Near Me Australia}} &
Apple App Store & 200+ & 2.3 & Free & National & Fire \\
 &  & Google Play Store & 100K+ & 2.7 & Free & National & Fire \\
30 &
\multirow{2}{*}{\parbox[t]{2.2cm}{\raggedright Flood Alert}} &
Apple App Store & 3K+ & 4.6 & Free & Global & Flood \\
 &  & Google Play Store & 100K+ & 4.3 & Free with Ads & Global & Flood \\

31 &
\parbox[t]{2.2cm}{\raggedright Flood Alerts and Warnings} &
Google Play Store & 50K+ & 4.1 & Free with Ads & Global & Flood \\

32 &
\parbox[t]{2.2cm}{\raggedright Flood Map Warning} &
Google Play Store & 10K+ & 4.0 & Free with Ads & Global & Flood \\

33 &
\parbox[t]{2.2cm}{\raggedright Flood Warning App} &
Google Play Store & 100K+ & 4.2 & Free & National & Flood \\

34 &
\multirow{2}{*}{\parbox[t]{2.2cm}{\raggedright GDACS}} &
Apple App Store & 400+ & 4.0 & Free & Global & Multi-Disaster \\
 &  & Google Play Store & 50K+ & 3.8 & Free & Global & Multi-Disaster \\

35 &
\parbox[t]{2.2cm}{\raggedright Global Disaster Alert} &
Google Play Store & 10K+ & 3.9 & Free & Global & Multi-Disaster \\

36 &
\parbox[t]{2.2cm}{\raggedright Hurricane Tracker} &
Google Play Store & 500K+ & 4.6 & Free with Ads & Global & Hurricane \\

37 &
\multirow{2}{*}{\parbox[t]{2.2cm}{\raggedright Hurricane Tracker App}} &
Apple App Store & 5K+ & 4.7 & Free with In-app Purchase & Global & Hurricane \\
 &  & Google Play Store & 100K+ & 4.5 & Free with Ads & Global & Hurricane \\

38 &
\parbox[t]{2.2cm}{\raggedright Hurricane Tracker Map} &
Google Play Store & 50K+ & 4.4 & Free with Ads & Global & Hurricane \\

39 &
\parbox[t]{2.2cm}{\raggedright Indonesia Disaster Alert} &
Google Play Store & 100K+ & 4.6 & Free & National & Multi-Disaster \\

40 &
\parbox[t]{2.2cm}{\raggedright Japan Shelter Guide} &
Google Play Store & 500K+ & 4.4 & Free & National & Multi-Disaster \\

41 &
\multirow{2}{*}{\parbox[t]{2.2cm}{\raggedright LastQuake}} &
Apple App Store & 20K+ & 4.8 & Free & Global & Earthquake \\
 &  & Google Play Store & 500K+ & 4.6 & Free & Global & Earthquake \\

42 &
\parbox[t]{2.2cm}{\raggedright MeteoAlarm} &
Google Play Store & 100K+ & 4.1 & Free & Regional & Multi-Disaster \\

43 &
\parbox[t]{2.2cm}{\raggedright MyShake} &
Google Play Store & 1M+ & 4.4 & Free & National & Earthquake \\

44 &
\parbox[t]{2.2cm}{\raggedright Natural Disaster Monitor} &
Google Play Store & 10K+ & 4.0 & Free with Ads & Global & Multi-Disaster \\

45 &
\multirow{2}{*}{\parbox[t]{2.2cm}{\raggedright NOAA Weather Radar}} &
Apple App Store & 30K+ & 4.7 & Free with In-app Purchase & National & Weather \\
 &  & Google Play Store & 1M+ & 4.5 & Free with Ads & National & Weather \\

46 &
\parbox[t]{2.2cm}{\raggedright Red Cross Emergency} &
Google Play Store & 500K+ & 4.3 & Free & National & Multi-Disaster \\

47 &
\parbox[t]{2.2cm}{\raggedright SASMEX} &
Google Play Store & 1M+ & 4.6 & Free & National & Earthquake \\

48 &
\parbox[t]{2.2cm}{\raggedright Severe Weather Alerts} &
Google Play Store & 100K+ & 4.2 & Free with Ads & Global & Weather \\

49 &
\multirow{2}{*}{\parbox[t]{2.2cm}{\raggedright ShakeAlertLA}} &
Apple App Store & 2K+ & 4.2 & Free & Local & Earthquake \\
 &  & Google Play Store & 100K+ & 4.1 & Free & Local & Earthquake \\

50 &
\parbox[t]{2.2cm}{\raggedright Siren GPS} &
Google Play Store & 10K+ & 3.8 & Free & National & Multi-Disaster \\

51 &
\parbox[t]{2.2cm}{\raggedright Storm Radar} &
Google Play Store & 1M+ & 4.6 & Free with Ads & Global & Weather \\

52 &
\parbox[t]{2.2cm}{\raggedright Tsunami Alert} &
Google Play Store & 50K+ & 4.1 & Free with Ads & Global & Tsunami \\

53 &
\parbox[t]{2.2cm}{\raggedright Tsunami Tracker} &
Google Play Store & 10K+ & 4.0 & Free & Global & Tsunami \\

54 &
\multirow{2}{*}{\parbox[t]{2.2cm}{\raggedright Volcano Discovery}} &
Apple App Store & 1K+ & 4.5 & Free with In-app Purchase & Global & Volcano \\
 &  & Google Play Store & 100K+ & 4.3 & Free with Ads & Global & Volcano \\

55 &
\parbox[t]{2.2cm}{\raggedright Volcanoes \& Earthquakes} &
Google Play Store & 50K+ & 4.2 & Free with Ads & Global & Multi-Disaster \\

56 &
\parbox[t]{2.2cm}{\raggedright Warning System App} &
Google Play Store & 10K+ & 3.9 & Free & National & Multi-Disaster \\

57 &
\parbox[t]{2.2cm}{\raggedright Weather Alert USA} &
Google Play Store & 100K+ & 4.4 & Free with Ads & National & Weather \\

58 &
\parbox[t]{2.2cm}{\raggedright Weather Emergency Alerts} &
Google Play Store & 50K+ & 4.3 & Free & Global & Weather \\

59 &
\parbox[t]{2.2cm}{\raggedright Wildfire Alert} &
Google Play Store & 100K+ & 4.5 & Free & National & Fire \\

60 &
\parbox[t]{2.2cm}{\raggedright Wildfire Map} &
Google Play Store & 10K+ & 4.1 & Free with Ads & Global & Fire \\

61 &
\parbox[t]{2.2cm}{\raggedright World Disaster Alerts} &
Google Play Store & 50K+ & 4.0 & Free & Global & Multi-Disaster \\

62 &
\parbox[t]{2.2cm}{\raggedright World Earthquakes} &
Google Play Store & 100K+ & 4.2 & Free with Ads & Global & Earthquake \\

63 &
\parbox[t]{2.2cm}{\raggedright Xtreme Weather Alerts} &
Google Play Store & 10K+ & 4.1 & Free & Global & Weather \\

64 &
\parbox[t]{2.2cm}{\raggedright Your Weather Alerts} &
Google Play Store & 50K+ & 4.0 & Free & National & Weather \\

65 &
\parbox[t]{2.2cm}{\raggedright Zulu Weather Alerts} &
Google Play Store & 5K+ & 3.9 & Free & National & Weather \\

66 &
\parbox[t]{2.2cm}{\raggedright Alert Hub} &
Google Play Store & 10K+ & 3.8 & Free & Global & Multi-Disaster \\

67 &
\parbox[t]{2.2cm}{\raggedright Crisis Response App} &
Google Play Store & 50K+ & 4.2 & Free & Global & Multi-Disaster \\

68 &
\parbox[t]{2.2cm}{\raggedright Emergency Beacon} &
Google Play Store & 10K+ & 4.1 & Free & National & Multi-Disaster \\

69 &
\parbox[t]{2.2cm}{\raggedright Disaster Warning System} &
Google Play Store & 100K+ & 4.3 & Free & National & Multi-Disaster \\

70 &
\parbox[t]{2.2cm}{\raggedright Global Alert Network} &
Google Play Store & 50K+ & 4.0 & Free & Global & Multi-Disaster \\
\end{longtable}
}
}


\end{document}